\newcommand{\be}[0]{\begin{equation}}
\newcommand{\ee}[0]{\end{equation}}
\newcommand{\ba}[0]{\begin{eqnarray}}
\newcommand{\ea}[0]{\end{eqnarray}}    

\documentstyle[12pt]{article}

\begin{document}
\Large
\hfill\vbox{\hbox{IPPP/01/11}\hbox{DCPT/01/22}\hbox{March 2001}}
\nopagebreak
\vspace{0.75cm}
\begin{center}
\LARGE
{\bf Renormalon-inspired
resummations for vector and scalar correlators- estimating the uncertainty 
in ${\alpha}_{s}({{M^2_\tau}})$ and
${\alpha}({{M^2_Z}})$}

\vspace{0.6cm}
\Large
C.J. Maxwell$^{1)}$ and  A. Mirjalili$^{2)}$
\vspace{0.4cm}
\large
\begin{em}

Centre for Particle Theory, University of Durham\\
South Road, Durham, DH1 3LE, England
\end{em}
\vspace{1cm}
\end{center}
\normalsize
\vspace{0.45cm}
\centerline{\bf Abstract}
\vspace{0.3cm}
We perform an all-orders resummation of the QCD Adler $D$-function for
the vector correlator, in which the portion of perturbative coefficients containing the
leading power of $b$, the first beta-function coefficient, is resummed. 
To avoid a renormalization scale dependence when we match the
resummation to the exactly known next-to-leading order (NLO), and next-NLO (NNLO) results,
we employ the Complete Renormalization Group Improvement (CORGI) approach
in which all RG-predictable ultra-violet logarithms are resummed to all-orders,
removing all dependence on the renormalization scale. We can also  obtain fixed-order CORGI results.
Including suitable weight-functions we can numerically integrate
these results for the $D$-function in the complex energy plane to obtain
so-called ``contour-improved'' results for the ratio $R$ and
its tau decay analogue ${R}_{\tau}$. We use the difference between the
all-orders and fixed-order (NNLO) results to estimate the uncertainty in
${\alpha}_{s}({M^2_Z})$ extracted from ${R}_{\tau}$ measurements, and find
${\alpha}_{s}({M^2_Z})=0.120{\pm}0.002$. We also estimate the corresponding
uncertainty in ${\alpha}({M^2_Z})$ arising from hadronic corrections by
considering the uncertainty in ${R(s)}$, in the low-energy
region, and compare with other estimates. Analogous resummations are also given for
the scalar correlator. As an adjunct to these studies we show how fixed-order
contour-improved results can be obtained analytically in closed form at the
two-loop level in terms of the Lambert $W$-function and hypergeometric functions.
\vfill\footnoterule\noindent
$^1$) C.J.Maxwell@durham.ac.uk\\
$^2$) Abolfazl.Mirjalili@durham.ac.uk
\
\newpage
\section*{1 Introduction}
The correlator of two vector currents in the Euclidean region is a fundamental
ingredient in constructing a number of inclusive hadronic QCD observables
of great importance in testing the theory. By taking a logarithmic energy
derivative one can define the so-called Adler $D$-function, $D(s)$. By analytical
continuation to the Minkowski region this quantity can then be directly related
to the ratio ,${R}(s)$, of the total ${e}^{+}{e}^{-}$ hadronic cross section to the
point leptonic cross-section, and also to the analogous ratio ${R}_{\tau}$ of the
total hadronic decay width of the $\tau$ lepton normalized to the leptonic
decay width. The analytical continuation can be elegantly formulated as a 
contour integration of $D(s)$ together with a weight function around a circle
in the complex energy $s$-plane \cite{r1,r2}. Performing this integration numerically
with $D(s)$ approximated at some fixed order of perturbation theory then automatically
resums to all-orders an infinite subset of potentially large analytical continuation
terms involving powers of ${\pi}^{2}$ and beta-fuction coefficients, which arise
in the running of the coupling around the integration contour. These terms are
usually truncated in the direct fixed-order perturbative expansion of the
Minkowskian quantity. Such approximations are referred to as ``contour-improved''.\\

The remaining uncertainty in these ``contour-improved'' predictions for $R$
and ${R}_{\tau}$ comes from the uncalculated higher order terms in the perturbation
series for $D(s)$, which has presently only been computed exactly to O$({{\alpha}_{s}}^{3})$ in
the limit of massless quarks \cite{r3}, and with some approximations including the contribution
of top and bottom quarks \cite{r4}. We shall assume massless quarks in these investigations.
There are also effects due to non-perturbative power
corrections \cite{r2,r5,r6}. We shall focus in this work on the former perturbative uncertainties.
There are two interrelated aspects to these. Fixed order perturbation theory predictions
have a dependence on the renormalization scheme (RS) chosen to define the coupling. In
particular they depend on a dimensionful renormalization scale ${\mu}$. Further, given a choice of RS there 
is an uncertainty due to the unknown O$({\alpha}_{s}^{4})$ and higher uncalculated
perturbative terms. To estimate this one needs to perform a necessarily approximate
all-orders resummation of these terms. A well-motivated framework to accomplish this
is provided by the so-called ``leading-$b$'' approximation \cite{r7,r8} (sometimes also referred to as
``naive nonabelianization'' [9-11]), which amounts to resumming to all-orders the portion
of perturbative coefficients containing the highest power of $b=\frac{1}{6}(11N-2{N}_{f})$
, the first beta-function coefficient for SU($N$) QCD with $N_f$ active massless quark flavours. 
This can be accomplished since in the large-$N_f$ limit one has an exact all-orders result
for the Adler $D$-function [12-14]. The leading-$b$ resummation is then performed by replacing
$N_f$ by $(\frac{11}{2}N-3b)$. Whilst the leading-$b$ approach is motivated by the structure
of renormalon singularities in the Borel plane \cite{r7,r8,r10,r11}, and also by a QCD skeleton
expansion \cite{r15}, it is effectively just the first ``one chain'' term in the skeleton expansion, and
 does not include the multiple exchanges of renormalon chains
needed to build the full asymptotic behaviour of the perturbative coefficients, and there
are no firm guarantees as to its accuracy. The strongest statement that can be made
follows from an analysis of the operators which build the leading ultraviolet
renormalon singularity \cite{r16,r17}. One can prove that in the case of the vector Adler function
the re-expansion of the leading-$b$ result in powers of $N_f$ correctly gives
the asymptotics of the portion of perturbative coefficients proportional to
${N_f}^{n-r}{N}^{r}$ in ${n}^{\rm{th}}$-order perturbation theory, with
accuracy O($1/n$) \cite{r17}. In practice for the exact NLO and NNLO coefficients of
the Adler $D$-function the level of agreement of these individual coefficients 
is at the ten percent level, much better than
would be expected from the above weak asymptotic result, remarkably in the ${N_f}=0$
or large-$N$ limit agreement is at the few percent level \cite{r8}. We may therefore hope that
the leading-$b$ approximation is indicative of the size of uncalculated higher-order
corrections. A remaining difficulty, first emphasised in \cite{r18}, is the scale dependence
of the leading-$b$ resummations, if one tries to match them to the known {\it exact}
NLO and NNLO perturbative coefficients. This matching ambiguity means that {\it any}
result may be obtained by varying the scale . It was pointed out that claims \cite{r5,r11} that
comparison of fixed-order and ``leading-$b$'' resummed results indicated rather
large uncalculated perturbative corrections for ${R}_{\tau}$ were undermined by
this matching problem. In Ref.\cite{r18,r19} the difficulty was resolved by performing
a leading-$b$ resummation for the Effective Charge beta-function \cite{r20} corresponding
to $D(s)$. This scheme-invariant construct then unambiguouly determines ${R}_{\tau}$.
This approach revealed a rather small uncertainty due to uncalculated higher-order
corrections. \\

In this paper we wish to formulate the resummations in a very closely related, 
but technically much more straightforward way. The renormalization scale, ${\mu}$,
dependence of fixed-order QCD perturbation theory is an artefact of the way 
renormalization group (RG) improvement is customarily performed. The two
crucial features are the use of a scale ${\mu}$ proportional to the physical
energy scale, $Q$, of the process, and the truncation of the perturbation series
at fixed-order. As argued recently  \cite{r21} one should rather
keep $\mu$ {\it strictly independent} of $Q$. Fixed-order perturbation theory
with $\mu$ constant does not then satisfy asymptotic freedom, and one is forced
to sum to all-orders the RG-predictable unphysical logarithms of ${\mu}$ and
physical ultraviolet (UV) logarithms of $Q$ from which the perturbative coefficients
are built. This so-called ``Complete RG-improvement'' (CORGI) \cite{r21} serves to
cancel all ${\mu}$-dependence between the  unphysical renormalised coupling
${\alpha}_{s}({\mu})$, and the unphysical logarithms of $\mu$ in the coefficients,
and one directly trades unphysical ${\mu}$ dependence for the physical $Q$-dependence.
The idea can also be generalized to processes, such as structure function moments,
which involve a factorization scale as well as a renormalization scale \cite{r21}.
The CORGI approach as formulated in Ref.\cite{r21} is exactly equivalent at NLO to the
Effective Charge approach of Grunberg \cite{r20} in which the UV logarithms are also
completely resummed in exactly the same way, whilst the remaining
RG-predictable effects are parametrized in a different, but {\it a priori}
equally reasonable way. Our plan is to perform a leading-$b$ resummation
for the Adler-$D$ function in the CORGI approach. As we shall see this is
extremely straightforward to implement and the resulting resummed result can
be written as a sum over exponential integral functions, representing the
contributions of the ultraviolet and infra-red renormalons in the Borel plane.
In contrast the resummation of the Effective Charge beta-function in Refs.\cite{r18,r19} involved
a complicated numerical inversion of a function. We anticipate that the
two approaches should yield very similar results. We shall obtain leading-$b$ resummed
and contour-improved CORGI results for the quantities ${R}(s)$, and
${R}_{\tau}$. As in Refs.\cite{r18,r19} we shall use the difference in fixed-order and
resummed results to estimate the uncertainty in ${\alpha}_{s}({M^2_Z})$ obtained
from measurements of ${R}_{\tau}$, using more recent experimental data \cite{r22}. We
shall also fit the leading-$b$ resummation to the spectral distribution for
hadronic ${\tau}$ decay \cite{r22,r23}. We shall attempt to estimate the uncertainty in the
hadronic corrections to the value of the QED coupling at the $Z$ pole,
 ${\alpha}({M^2_Z})$, using the resummed and fixed-order results for ${R}(s)$
in the energy ranges $5<\sqrt{s}<{\infty}$ GeV, and $2.8<\sqrt{s}<3.74$ GeV, and
using inclusive data in the remaining ranges, as in Ref.\cite{r24}. We shall compare our result
for ${\alpha}({M}_{Z}^{2})$ with that of Ref.\cite{r24}, which uses standard fixed-order
perturbation theory. Using recent large-$N_f$ results on the scalar correlator \cite{r24a}
we shall also perform a contour-improved leading-$b$ resummation for the Higgs
decay width. Finally, we show how the analytical continuation to the Minkowski
region can be performed in closed analytical form at the two-loop level in
terms of the Lambert $W$-function and hypergeometric functions.\\

The organization of the paper is as follows. In Section 2 we shall review the
contour integral representation of ${R}(s)$ and ${R}_{\tau}$ in
terms of $D(s)$, and describe a simple numerical algorithm for evaluating it.
In Section 3 we shall review the fixed-order perturbative, and all-orders
large-$N_f$ results for $D(s)$ , and show how the leading-$b$ all-orders resummation
in the CORGI approach can be written in closed form as a sum of exponential 
integral functions. In Section 4 we estimate the uncertainty in ${\alpha}_{s}({M^2_Z})$
obtained from ${R}_{\tau}$ measurements, and also fit the leading-$b$ resummed
results to the spectral function. In Section 5 we shall estimate the uncertainty
in the hadronic corrections to ${\alpha}({M^2_Z})$ as discussed above. In Section 6
we perform a contour-improved leading-$b$ resummation for the Higgs decay width,
and in Section 7 discuss how the analytical continuation to the Minkowski region
can be performed analytically at the two-loop level in terms of the Lambert $W$-function
and hypergeometric functions. Section 8 contains a discussion and our conclusions.
\section*{2 Contour integral representation of Minkowski observables}
We shall mainly be concerned in this work with two inclusive QCD observables.
The first is the ${e}^{+}{e}^{-}$ $R$-ratio, defined by
\be
{R}{\equiv}\frac{{\sigma}({e}^{+}{e}^{-}\rightarrow{\rm{hadrons}})}{{\sigma}(
{e}^{+}{e}^{-}\rightarrow{\mu}^{+}{\mu}^{-})}\;.
\ee
In SU($N$) QCD perturbation theory
\be
{R}(s)=N{\sum_{f}}{Q_f}^{2}\left(1+\frac{3}{4}{C_F}\tilde{R}(s)\right)+
{\left({\sum_{f}}{Q_f}\right)}^{2}\bar{R}(s)\;,
\ee
with $Q_f$ denoting the quark charges, summed over the flavours accessible
at a given energy. $C_F$ is the SU($N$) Casimir ${C_F}=({N}^{2}-1)/2N$,
and $s$ denotes the timelike Minkowski squared momentum transfer. $\tilde{R}$
denotes the perturbative corections to the parton model result. It has the
perturbative expansion
\be
\tilde{R}(s)=a(1+{\sum_{n>0}}{r_n}{a}^n)\;,
\ee
where $a{\equiv}{\alpha}_{s}({\mu}^{2})/{\pi}$ denotes the RG-improved coupling.
$\bar{R}$ denotes so-called ``light-by-light'' contributions which enter at
 O$({a}^{3})$. We shall ignore this term in our all-orders resummations.
The ratio ${R}_{\tau}$ is defined analogously as a ratio of the total $\tau$
hadronic decay width to its leptonic decay width,
\be
{R}_{\tau}{\equiv}\frac{{\Gamma}({\tau}\rightarrow{{\nu}_{\tau}}+{\rm{hadrons}})}
{{\Gamma}({\tau}\rightarrow{{\nu}_{\tau}}{e}^{-}{\bar{\nu}}_{e})}\;.
\ee
Its perturbative expansion has the form
\be
{R}_{\tau}=N({|{V}_{ud}|}^{2}+{|V_{us}|}^{2}){S_{EW}}
\left[1+{\frac{5}{12}}{\frac{{\alpha}({m_{\tau}}^{2})}{\pi}}
+{\tilde{R}}_{\tau}+{\delta}_{PC}\right]\;,
\ee
with ${V}_{ud}$ and ${V}_{us}$ CKM mixing matrix elements. Since the energy scale
$s={{m}^2_{\tau}}$ lies below the threshold for charmed hadron production
only three flavours $u$,$d$,$s$, are active. The ${\alpha}({{m}^2_{\tau}})$ term
denotes the leading QED electromagnetic corrections, and ${S}_{EW}{\approx}1.0194$ \cite{r25}
represents further electroweak corrections. ${\delta}_{PC}$ denotes possible power
corrections. ${\tilde{R}}_{\tau}$ has a perturbative
expansion
\be
{\tilde{R}}_{\tau}=a(1+{\sum_{n>0}}{r}^{\tau}_{n}{a}^{n})\;.
\ee
In this case there are no "light-by-light'' corrections because summing over $u$,$d$ and $s$
quarks ${({\sum{Q_f})}^{2}}$=0.
Both $R$ and ${R}_{\tau}$ can be directly expressed in terms of the transverse part
of the correlator of two vector currents in the Euclidean region,
\be
({q}_{\mu}{q}_{\nu}-{g}_{{\mu}{\nu}}{q}^{2}){\Pi}(s)=4{\pi}^2i{\int}{d}^{4}x{e}^{iq.x}
<0|T[{j}_{\mu}(x){j}_{\nu}(x)]|0>\;,
\ee
where $s=-{q}^{2}>0$. In fact it is convenient to take a logarithmic derivative
with respect to $s$ and define the Adler $D$-function,
\be
D(s)=-s\frac{d}{ds}{\Pi}(s)\;.
\ee
This can be represented by an expression analogous to Eq.(2) involving $\tilde{D}$ and
$\bar{D}$, where $\tilde{D}(s)$ has the perturbative expansion
\be
\tilde{D}(s)=a(1+{\sum_{n>0}}{d_n}{a}^{n})\;.
\ee
A generic Minkowskian observable $\hat{R}({s_0})$ can then be related to $\tilde{D}(-{s})$
by analytical continuation from Euclidean to Minkowski. This can be elegantly formulated
as an integration around a circular contour in the complex energy squared $s$-plane \cite{r19},
\be
{\hat{R}}({s}_{0})={\frac{1}{2\pi}}{\int_{-\pi}^{\pi}}{W({\theta})}{\tilde{D}}
({{s}_{0}}e^{i{\theta}})d{\theta}\;,
\ee
where ${W({\theta})}$ is a weight function which depends on the observable 
${\hat{R}}$. For $W({\theta})=1$ one has $\hat{R}({s_0})=\tilde{R}({s_0})$,
and for $W({\theta})=(1+2{e}^{i{\theta}}-2{e}^{3i{\theta}}-{e}^{4i{\theta}})$ one
has $\hat{R}({m}_{\tau}^{2})$=${\tilde{R}}_{\tau}$. 
If one expands $\tilde{D}({s_0}{e}^{i{\theta}})$ as a perturbation series in
$\bar{a}{\equiv}a({s_0}{e}^{i{\theta}})$ and numerically performs the $\theta$ integration
term-by-term one obtains ``contour-improved'' perturbative results in which at
each order an infinite subset of analytical continuation terms present in
the conventional perturbation series of Eqs.(3),(6) are resummed. These terms
are potentially large and involve powers of ${\pi}^{2}$ and beta-function
coefficients, as is easily seen by expanding ${\bar{a}}$ in powers of $a({s_0})$
and integrating. In this paper we shall focus on this ``contour-improved''
version of perturbation theory. In Ref.\cite{r19} detailed comparisons of the performance
of the two versions were made, and the importance of resumming the analytical
continuation terms was emphasised.\\

An obvious numerical algorithm for evaluating the integral in Eq.(10) is to split 
the range from ${\theta}$=0,${\pi}$ into $K$ steps of size 
${\Delta}{\theta}={\pi}/K$ and perform a sum over
the integrand evaluated at
${{\theta}_n}$=n${\Delta}{\theta}$, n=0,1,...,K. So 
that
\be
{\hat{R}}({s}_{0}){\simeq}{\frac{{\Delta}{\theta}}{2{\pi}}}[{W({0})}{\tilde{D}}
({{s}_{0}})
+2Re{\sum_{n=1}^{K}}{W({\theta}_n)}{\tilde{D}}({{s}_{n}})]
\ee
where ${s_n}{\equiv}{s_{0}}{e^{in{\Delta}{\theta}}}$. In practice we perform a 
Simpson's Rule evaluation. Writing the perturbation expansion for ${\tilde{D}}({s_n})$
we have 
\be
{\tilde{D}}({s_n})={\bar{a}}_{n}+{d_1}{{\bar{a}}_{n}}^{2}+{d_2}{{\bar{a}}_{n}}^{3}+\ldots\;,
\ee
where we have defined ${\bar{a}}_{n}{\equiv}a({s_n})$. An efficient strategy \cite{r26} is to
start with ${\bar{a}}_{0}=a({s_0})$ and use Taylor's theorem step-by-step to evolve
${\bar{a}}_{n}$ to ${\bar{a}}_{n+1}$, using
\ba
\setlength\arraycolsep{2pt}
{{\bar{a}}_{n+1}}&=&{{\bar{a}}_{n}}-i{\frac{{\Delta}{\theta}}{2}}b{B({\bar{a}}_{n})}
-{\frac{{\Delta}{\theta}^2}{8}}b^2{B({\bar{a}}_{n})}{{B}{'}}({\bar{a}}_{n})
+i{\frac{{\Delta}{\theta}^3}{48}}b^3[{B({\bar{a}}_{n})}{{B{'}({\bar{a}}_{n})}}^2
\nonumber\\
&&+{B({\bar{a}}_{n})}^2{{B''({\bar{a}}_{n})}}]+O({\Delta}{\theta}^4)+...
\ea
where ${B(x)}$ is the truncated beta-function 
\be
{B(x)}=x^2+c{x}^{3}+{c_2}{x^4}+\dots\;,
\ee
so that ${\bar{a}}$ satisfies
\be
{\frac{\partial{\bar{a}}}{\partial{lns}}}=-{\frac{b}{2}}({\bar{a}^2}+c{\bar{a}^3}
+{c_2}{\bar{a}^4}+...)=-{\frac{b}{2}}B({\bar{a}})\;.
\ee
Here $b=(33-2{N_f})/6$, and $c=(153-19{N_f})/12b$ are the first two universal beta-function
coefficients for SU($3$) QCD with $N_f$ active massless quark flavours. The higher coefficients
${c_i}, i>1$ are scheme-dependent. The above use of Taylor's theorem is much faster to
implement than the standard approach \cite{r5} of solving the integrated beta-function equation with
complex renormalization scale $s_n$ to find ${\bar{a}}_{n}$ at each step.
\section*{3 Fixed-order and resummed expressions for $D(s)$ in the CORGI approach}
In the CORGI approach one avoids renormalization scale ${\mu}$-dependence by performing
a complete resummation of the ultraviolet logarithms which build the dependence of
the observable on the physical energy scale \cite{r21}. This is equivalent to directly
relating the observable to the dimensional transmutation parameter of the theory \cite{r27}, ${\Lambda}_{\overline{MS}}$
say. In this way one can write the CORGI series for ${\tilde{D}}(s)$,
\be
{\tilde{D}}(s)= {a}_{0}(s)+{X}_{2}{{a}^3_{0}(s)}+{X}_{3}{{a}^4_{0}}+\ldots+{X}_{n}{{a}^{n+1}_{0}}+\ldots\;.
\ee 
Here ${a}_{0}(s)$ is the CORGI coupling which may be written in terms of the Lambert $W$-function
defined implicitly by $W(z){\exp}(W(z))=z$ \cite{r28} as,
\ba
{a}_{0}(s)&=&-\frac{1}{c[1+W(z(s))]}
\nonumber \\
z(s)&{\equiv}&-\frac{1}{e}{\left(\frac{\sqrt{s}}{{\Lambda}_{D}}\right)}^{-b/c}\;,
\ea
where ${\Lambda}_{D}{\equiv}{e}^{d/b}{(2c/b)}^{-c/b}{\Lambda}_{\overline{MS}}$, with $d$
 the NLO perturbative coefficient $d_1$ for ${\tilde{D}}(s)$ in Eq.(9), in the $\overline{MS}$
scheme with ${\mu}^{2}=s$. 
${a}_{0}(s)$ is the coupling in the scheme with ${\mu}^{2}={e}^{-2d/b}s$,
and the non-universal beta-function coefficients, ${c}_{i},\;(i>1)$ all zero. In this scheme ${d}_{1}=0$,
and it is exactly
equivalent at NLO to the Effective Charge approach of Grunberg \cite{r20}. Standard RG-improvement
in this scheme completely resums all ultraviolet logarithms, and is equivalent to the
CORGI approach which can be formulated in {\it any scheme} \cite{r21}.
${X_2}$ is the NNLO scheme-invariant combination
\be 
{X_2}={c_2}+{d_2}-{c}{d_1}-{d}_{1}^{2}\;,
\ee
built from the perturbative coefficients ${d_1}$ and ${d_2}$ and beta-function coefficients.
The NLO and NNLO coefficients $d_1$ and $d_2$ are known exactly \cite{r3} and so NNLO
contour-improved CORGI predictions can be straightforwardly obtained for Minkowski
observables ${\hat{R}}({s_0})$, using the numerical integration described in Section 2.
Since ${a}_{0}(s)$ is known in closed form in terms of the Lambert $W$-function, which has
a well-defined branch structure in the complex plane, one can evaluate it directly, avoiding
the Taylor's theorem trick in Eq.(13). In fact one needs to use the ${W}_{-1}$ branch of the
function (in the nomenclature of Ref.\cite{r28}) on the range of integration [$0,{\pi}$], and the
$W_1$ branch on the range [$-{\pi},0$]. As we shall discuss in Section 7 one can, in fact,
avoid using the numerical Simpson's Rule integration all together for the case of the
${e}^{+}{e}^{-}$ $R$-ratio where $W({\theta})=1$, and perform the integral in closed form
in terms of logarithms of the $W$-function.\\

In order to assess the likely accuracy of the fixed-order perturbative approximation we
can attempt to approximate at the present uncalculated coefficients ${d_i}$, $(i>2)$ in
${\tilde{D}}(s)$ using the so-called ``leading-$b$'' approximation. A given coefficient
$d_n$ can be written as an expansion in powers of $N_f$, so that we have
\be
{d_n}= {d}_{n}^{[n]}{N}_{f}^{n}+{d}_{n}^{[n-1]}{N}_{f}^{n-1}+{\ldots}+{d}_{n}^{[0]}\;.
\ee
The large-$N_f$ coefficient ${d}_{n}^{[n]}$ can be computed exactly to all-orders since it
derives from a restricted set of diagrams in which a chain of $n$ fermion bubbles is
inserted in the initiating quark loop \cite{r12,r13}. Motivated by the structure of renormalon
singularities in the Borel plane one can convert this expansion into the so-called
leading-$b$ expansion \cite{r7,r8}, by substituting ${N_f}=(33/2-3b)$, to obtain
\be
{d_n}={d}_{n}^{(n)}{b}^{n}+{d}_{n}^{(n-1)}{b}^{n-1}+{\ldots}+{d}_{n}^{(0)}\;.
\ee
The leading-$b$ term ${d}_{n}^{(L)}{\equiv}{d}_{n}^{(n)}{b}^{n}$ is then used to
approximate ${d_n}$. Since ${d}_{n}^{(L)}={(-3)}^{n}{d}_{n}^{[n]}{b}^{n}$ it is
known to all-orders. Using the exact large-${N_f}$ result one
finds that
the explicit all-orders result
for ${d}_{n}^{(L)}$ in the so-called $V$-scheme, i.e. ${\overline{MS}}$ with
scale ${\mu}^{2}={e}^{-5/3}s$, is given by \cite{r13}
\ba
{{d}^{(L)}_{n}}(V)&=&{\frac{-2}{3}}{n!}{\frac{(n+1)}{2^n}}[-2n-{\frac{n+6}{2^{n+2}}}
\nonumber \\
&&+{\frac{16}{n+1}}
{\sum_{{{\frac{n}{2}}+1}{>}s{>}0}}{s(1-2^{-2s})(1-2^{2s-n-2}}
{\zeta}_{2s+1})]{b}^{n}\;.
\ea
The resulting leading-$b$ resummation
\be
{\tilde{D}}^{(L)}=a(1+{\sum_{k=0}^{{\infty}}}{d}_{k}^{(L)}{a}^{k}),
\ee
may then be defined as a principal value (PV) regulated Borel Sum,
\be
{\tilde{D}}^{(L)}(1/a)=PV{\int_{0}^{\infty}}{dz}{e}^{-z/a}B[{\tilde{D}}^{(L)}](z)\;.
\ee
Here $B[{\tilde{D}}^{(L)}](z)$, denotes the Borel transform which contains an infinite
set of single and double poles at $z={z_l}=\frac{2l}{b}$ corresponding to
infra-red renormalons , ${IR}_{l}$, and an infinite set of ultra-violet renormalons, ${UV}_{l}$,
at $z=-{z_l}$.
The structure is
\be
B[{\tilde{D}}^{(L)}](z)={\sum_{j=1}^{\infty}}\frac{{A}_{0}(j)+{A}_{1}(j)z}{{(1+\frac{z}{{z_j}})}^{2}}+
\frac{{B}_{0}(2)}{(1-\frac{z}{{z_2}})}+
{\sum_{j=3}^{\infty}}\frac{{B}_{0}(j)+{B}_{1}(j)z}{{(1-\frac{z}{{z_j}})}^{2}}\;.
\ee
The residues at these poles can be computed from the exact
all-orders large-$N_f$ result. The $UV$ and $IR$ renormalon contributions
can then be easily expressed in terms of the exponential integral function,
\be
Ei(x)=-{\int_{-x}^{\infty}}{dt}\frac{{e}^{-t}}{t}\;,
\ee
where for $IR$ renormalons $x>0$ and one defines $Ei(x)$ by taking the Cauchy
principal value of the integral. The arbitrariness in regulating the
$IR$ renormalon contributions reflects the fact that the perturbation series
needs to be combined with the power corrections of the operator product
expansion (OPE) to obtain a well-defined result \cite{r29}.
The absence of a relevant operator
of dimension two in the OPE for the vector correlator is in accord with the
fact that the singularity ${IR}_{1}$ is not present, and the nearest singularity
to the origin in the Borel plane is in fact ${UV}_{1}$, which generates the leading 
asymptotic behaviour \cite{r8},
\be
{d}_{n}^{(L)}(V){\approx}\frac{(12n+22)}{27}{n!}{\left(-\frac{1}{2}\right)}^{n}{b}^{n}\;.
\ee
One can then write the $UV$ renormalon and $IR$ renormalon contributions as infinite
sums over the $Ei$ functions,
\ba
{\tilde{D}}^{L}(F){{\mid}_{UV}}&=&{\sum_{j=1}^{\infty}}{z_j}\{e^{F(a){z_j}}Ei(-F{z_j})
[F{z_j}({{A_0}(j)}-{z_j}{{A_1}(j)})-{z_j}{{A_1}(j)}]
\nonumber\\
&&+({{A_0}(j)}-{z_j}{{A_1}(j)}))\}\;,
\ea
and
\ba 
{\tilde{D}}^{L}(F){{\mid}_{IR}}&=&{e^{-F{z_2}}}{z_2}{{B_{0}}(2)}Ei(F{z_2})
\nonumber\\
&&{\sum_{j=3}^{\infty}}
{z_j}\{e^{-F{z_j}}Ei(F{z_j})
[F{z_j}({{B_0}(j)}+{z_j}{{B_1}(j)})-{z_j}{{B_1}(j)}]
\nonumber\\
&&-({{B_0}(j)}+{z_j}{{B_1}(j)})\}\;.
\ea
Here we have defined ${F{\equiv}1/{a_V}}$, where ${a_V}$ is the coupling in the V-scheme.
The ${{A_0}(j),{A_1}(j)}$ are related to the residues of the ${UV}_{j}$ poles, with \cite{r8}
\be 
{A_0}(j)=\frac{8}{3}\frac{{(-1)}^{j+1}(3{j}^{2}+6j+2)}{{j}^{2}{(j+1)}^{2}{(j+2)}^{2}}\;\;\;
{A_1}(j)=\frac{4}{3}\frac{b{{(-1)}^{j+1}(2j+3)}}{{j}^{2}{(j+1)}^{2}{(j+2)}^{2}}\;.
\ee
Because of the conformal symmetry \cite{r30} of the vector correlator the ${UV}$ residues
are directly related to the ${IR}$ residues with ${{B_0}(j)=-{A_0}(-j)}$ and {\hspace{.1cm}${{B_1}(j)}$=}
\newline
${-{A_1}(-j)}$ for ${j{>}2}$, and ${{B_0}(1)={B_1}(1)={B_1}(2)=0}$, and
${{B_0}(2)=1}$ \cite{r8}. 
To evaluate the contour integral in the
complex ${s}$-plane using this ${{\tilde{D}}^{(L)}(F)}$ result one needs to
modify the definition of the ${Ei}$ functions to cope with the fact that their
argument involves ${1/{a}_{V}({s_0}{e}^{i{\theta}})}$ which is complex for
nonzero ${\theta}$. The appropriate generalization uses the function ${Ei(n,z)}$
defined by
\be
Ei(n,z)={\int_{1}^{\infty}}{dt}\frac{{e}^{-tz}}{{t}^{n}}\;.
\ee
This function is analytic everywhere in the cut complex $z$-plane, but has
a branch cut along the negative real axis. One needs to replace ${Ei(-F{z}_{j})}$
in the ${UV}$ contribution by ${-Ei(1,F{z}_{j})}$, and ${Ei(F{z}_{j})}$ in the ${IR}$
contribution by ${-Ei(1,-F{z}_{j})+{i}{\pi}sign(Im(F{z}_{j}))}$, where the discontinuity
across the branch cut is removed by the final ${i{\pi}}$ contribution \cite{r8}. The
final result for ${{\tilde{D}}(F)}$ is simply the sum of the ${UV}$ and ${IR}$ 
contributions. The sums in Eqs.(27),(28) are rapidly convergent since the
${A(j)}$ and ${B(j)}$ coefficients have a ${{j}^{-4}}$ dependence for large ${j}$.
For the numerical results to be reported in Section 4 we used ${{N}_{UV}=15}$
and ${{N}_{IR}=17}$ terms respectively in the two sums. It is sensible to
arrange that ${{N}_{IR}={N}_{UV}+2}$ , since the symmetry properties above
mean that ${{A}_{0}(j)=-{B}_{0}(j+2)}$, this ensures that the O${(a)}$ term
in the perturbation series of Eq.(9) has the correct unit coefficient
${{B}_{0}(2)=1}$.\\

The final step is to use the above results to perform an all-orders resummation
in the CORGI approach. We would like to formally perform the resummation
\be
{\tilde{D}}_{CORGI}={a}_{0}+{X}_{2}{{a}^3_{0}}+{\sum_{n>2}}{X}_{n}^{(L)}{a}_{0}^{n+1}\;,
\ee
so that the exactly known NNLO $X_2$ coefficient is included , with the remaining unknown
coefficients approximated by ${X}_{3}^{(L)} ,{X}_{4}^{(L)},\ldots$, the leading-$b$
approximations. Note that ${a}_{0}$ is the full CORGI coupling defined in Eq.(17), so
that all the RG-predictable ultraviolet logarithms involving the exact NLO coefficient
${d}_{1}$ are completely resummed.
This  resummation is most easily achieved by noting that the combination \cite{r31} 
\be
{\rho}_{0}=b{ln}\left(\frac{\mu}{{\tilde{\Lambda}}}\right)-{d}_{1}({\mu})\;,
\ee
is scheme-independent. At the leading-$b$ level the coupling ${a}^{(L)}(s)$ is
defined by the one-loop formula
\be
{a}^{(L)}(s)=\frac{1}{b{\ln}({\sqrt{s}}/{\tilde{\Lambda}})}\;.
\ee
In the CORGI scheme in leading-$b$ approximation ${d}_{1}^{(L)}=0$, and so by
evaluating the invariant combination ${\rho_0}$ in Eq.(32) in the $V$ scheme and
the CORGI scheme one can relate the couplings in the two schemes,
\be
\frac{1}{{a}_{V}^{(L)}}=\frac{1}{{a}_{0}^{(L)}}+{d}_{1}^{L}(V)\;. 
\ee
It then follows straightforwardly that the formal resummation in Eq.(31) is given by
\be
{\tilde{D}}_{CORGI}={\tilde{D}}^{(L)}\left(\frac{1}{{a}_{0}}+{d}_{1}^{(L)}(V)\right)+({X}_{2}-{X}_{2}^{(L)})
{{a}^3_{0}}\;,
\ee
in which the ${\tilde{D}}^{(L)}$ term is the all-orders sum  with the exact ${X}_{2}$ replaced by
${X}_{2}^{(L)}$, and the second term corrects for this. One can obtain approximate
${\rm{N}}^{3}$LO and higher CORGI results by truncating the sum in Eq.(31). The ${X}_{n}^{(L)}$
can be readily calculated by using the leading-$b$ relation between the $V$-scheme and CORGI
couplings in (34). One easily finds
\be
{X}_{n}^{(L)}={\cal{C}}_{n+1}\left[{\sum_{k=1}^{\infty}}{d}_{n}^{(L)}(V)
{\left(\frac{a}{1+{a}{d}_{1}^{(L)}(V)}\right)}^{k+1}\right]\;,
\ee
where the symbol ${\cal{C}}_{n}[f(a)]$ denotes the coefficient of ${a}^{n}$ in the power
series expansion of $f(a)$. The ${d}_{n}^{(L)}(V)$ can be directly generated using the 
explicit result in Eq.(21).\\
Using the above results we can now straightforwardly generate all-orders resummed and fixed-order
contour-improved CORGI results for the Minkowski observables ${R}_{\tau}$ and $R$. We shall
perform some phenomenological studies in the next two sections.
\section*{4 Resummed versus fixed-order predictions for ${R}_{\tau}$}
The ratio ${R}_{\tau}$ defined by Eq.(4) has been the subject of a wide-ranging
experimental study by the ALEPH collaboration \cite{r22}. If events involving strange
quarks are removed from the data, they find ${R}_{\tau}=3.492{\pm}0.016$. Setting
${V}_{us}=0$, and ${V}_{ud}=0.9754{\pm}0.0007$, and {\hspace{.20cm}estimating} {\hspace{.20cm}the} {\hspace{.2cm}power} 
{\hspace{.21cm}correction}
\newpage
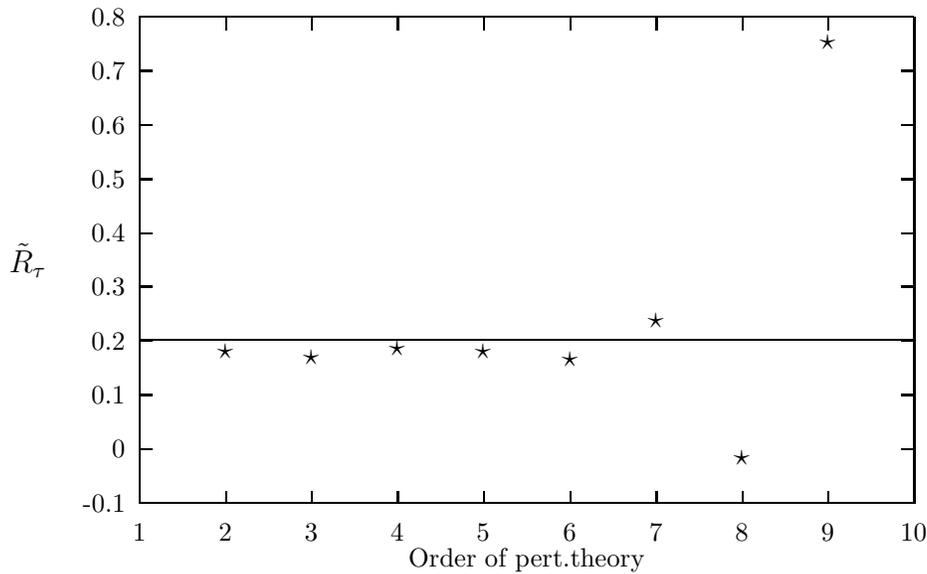
\begin{figure}
\begin{center}
\setlength{\unitlength}{0.240900pt}
\ifx\plotpoint\undefined\newsavebox{\plotpoint}\fi
\sbox{\plotpoint}{\rule[-0.200pt]{0.400pt}{0.400pt}}%
\begin{picture}(1500,900)(0,0)
\font\gnuplot=cmr10 at 10pt
\gnuplot
\sbox{\plotpoint}{\rule[-0.200pt]{0.400pt}{0.400pt}}%
\put(220.0,113.0){\rule[-0.200pt]{4.818pt}{0.400pt}}
\put(198,113){\makebox(0,0)[r]{-0.1}}
\put(1416.0,113.0){\rule[-0.200pt]{4.818pt}{0.400pt}}
\put(220.0,198.0){\rule[-0.200pt]{4.818pt}{0.400pt}}
\put(198,198){\makebox(0,0)[r]{0}}
\put(1416.0,198.0){\rule[-0.200pt]{4.818pt}{0.400pt}}
\put(220.0,283.0){\rule[-0.200pt]{4.818pt}{0.400pt}}
\put(198,283){\makebox(0,0)[r]{0.1}}
\put(1416.0,283.0){\rule[-0.200pt]{4.818pt}{0.400pt}}
\put(220.0,368.0){\rule[-0.200pt]{4.818pt}{0.400pt}}
\put(198,368){\makebox(0,0)[r]{0.2}}
\put(1416.0,368.0){\rule[-0.200pt]{4.818pt}{0.400pt}}
\put(220.0,453.0){\rule[-0.200pt]{4.818pt}{0.400pt}}
\put(198,453){\makebox(0,0)[r]{0.3}}
\put(1416.0,453.0){\rule[-0.200pt]{4.818pt}{0.400pt}}
\put(220.0,537.0){\rule[-0.200pt]{4.818pt}{0.400pt}}
\put(198,537){\makebox(0,0)[r]{0.4}}
\put(1416.0,537.0){\rule[-0.200pt]{4.818pt}{0.400pt}}
\put(220.0,622.0){\rule[-0.200pt]{4.818pt}{0.400pt}}
\put(198,622){\makebox(0,0)[r]{0.5}}
\put(1416.0,622.0){\rule[-0.200pt]{4.818pt}{0.400pt}}
\put(220.0,707.0){\rule[-0.200pt]{4.818pt}{0.400pt}}
\put(198,707){\makebox(0,0)[r]{0.6}}
\put(1416.0,707.0){\rule[-0.200pt]{4.818pt}{0.400pt}}
\put(220.0,792.0){\rule[-0.200pt]{4.818pt}{0.400pt}}
\put(198,792){\makebox(0,0)[r]{0.7}}
\put(1416.0,792.0){\rule[-0.200pt]{4.818pt}{0.400pt}}
\put(220.0,877.0){\rule[-0.200pt]{4.818pt}{0.400pt}}
\put(198,877){\makebox(0,0)[r]{0.8}}
\put(1416.0,877.0){\rule[-0.200pt]{4.818pt}{0.400pt}}
\put(220.0,113.0){\rule[-0.200pt]{0.400pt}{4.818pt}}
\put(220,68){\makebox(0,0){1}}
\put(220.0,857.0){\rule[-0.200pt]{0.400pt}{4.818pt}}
\put(355.0,113.0){\rule[-0.200pt]{0.400pt}{4.818pt}}
\put(355,68){\makebox(0,0){2}}
\put(355.0,857.0){\rule[-0.200pt]{0.400pt}{4.818pt}}
\put(490.0,113.0){\rule[-0.200pt]{0.400pt}{4.818pt}}
\put(490,68){\makebox(0,0){3}}
\put(490.0,857.0){\rule[-0.200pt]{0.400pt}{4.818pt}}
\put(625.0,113.0){\rule[-0.200pt]{0.400pt}{4.818pt}}
\put(625,68){\makebox(0,0){4}}
\put(625.0,857.0){\rule[-0.200pt]{0.400pt}{4.818pt}}
\put(760.0,113.0){\rule[-0.200pt]{0.400pt}{4.818pt}}
\put(760,68){\makebox(0,0){5}}
\put(760.0,857.0){\rule[-0.200pt]{0.400pt}{4.818pt}}
\put(896.0,113.0){\rule[-0.200pt]{0.400pt}{4.818pt}}
\put(896,68){\makebox(0,0){6}}
\put(896.0,857.0){\rule[-0.200pt]{0.400pt}{4.818pt}}
\put(1031.0,113.0){\rule[-0.200pt]{0.400pt}{4.818pt}}
\put(1031,68){\makebox(0,0){7}}
\put(1031.0,857.0){\rule[-0.200pt]{0.400pt}{4.818pt}}
\put(1166.0,113.0){\rule[-0.200pt]{0.400pt}{4.818pt}}
\put(1166,68){\makebox(0,0){8}}
\put(1166.0,857.0){\rule[-0.200pt]{0.400pt}{4.818pt}}
\put(1301.0,113.0){\rule[-0.200pt]{0.400pt}{4.818pt}}
\put(1301,68){\makebox(0,0){9}}
\put(1301.0,857.0){\rule[-0.200pt]{0.400pt}{4.818pt}}
\put(1436.0,113.0){\rule[-0.200pt]{0.400pt}{4.818pt}}
\put(1436,68){\makebox(0,0){10}}
\put(1436.0,857.0){\rule[-0.200pt]{0.400pt}{4.818pt}}
\put(220.0,113.0){\rule[-0.200pt]{292.934pt}{0.400pt}}
\put(1436.0,113.0){\rule[-0.200pt]{0.400pt}{184.048pt}}
\put(220.0,877.0){\rule[-0.200pt]{292.934pt}{0.400pt}}
\put(45,495){\makebox(0,0){${\tilde{R}_{\tau}}$}}
\put(828,23){\makebox(0,0){Order of pert.theory}}
\put(220.0,113.0){\rule[-0.200pt]{0.400pt}{184.048pt}}
\put(355,355){\raisebox{-.8pt}{\makebox(0,0){$\star$}}}
\put(490,346){\raisebox{-.8pt}{\makebox(0,0){$\star$}}}
\put(625,360){\raisebox{-.8pt}{\makebox(0,0){$\star$}}}
\put(760,355){\raisebox{-.8pt}{\makebox(0,0){$\star$}}}
\put(896,342){\raisebox{-.8pt}{\makebox(0,0){$\star$}}}
\put(1031,404){\raisebox{-.8pt}{\makebox(0,0){$\star$}}}
\put(1166,187){\raisebox{-.8pt}{\makebox(0,0){$\star$}}}
\put(1301,841){\raisebox{-.8pt}{\makebox(0,0){$\star$}}}
\put(220,370){\usebox{\plotpoint}}
\put(220.0,370.0){\rule[-0.200pt]{292.934pt}{0.400pt}}
\end{picture}
\caption{Fixed-order CORGI results 
for ${\tilde{R}_{\tau}}$ in ${\rm{N}}^{n}$LO perturbation theory (starred points), compared
to the all-orders resummation (solid line) fitted to ALEPH data.}
\label{fig:Figure1}
\end{center}
\end{figure}
{\hspace{-.7cm}contribution} to be ${\delta}_{PC}=-0.003{\pm}0.004$  \cite{r22}, one finds from Eq.(5) the
experimental value ${\tilde{R}}_{\tau}={0.2032}^{+0.0160}_{-0.0159}$. The QED
contribution has been neglected. One can then obtain all-orders leading-$b$ resummed and
fixed-order contour-improved CORGI results as described in Sections 2 and 3. We use
${N_f}=3$ and fix ${\Lambda}_{\overline{MS}}^{(3)}$ so that the all-orders result
reproduces the measured central value ${\tilde{R}}_{\tau}=0.203$. The results
are shown in Figure 1. The solid line is the all-orders resummed result fixed to
the data, and the starred points show the ${\rm{N}}^{n}$LO fixed-order CORGI results.
 We see that the NNLO ($n$=2) 
fixed-order result, which is the highest order exactly known,
is in rather good agreement with the all-orders resummation. The leading-$b$ approximated
${\rm{N}}^{n}$LO results show an oscillatory trend which becomes explosive for $n>7$,
where fixed-order perturbation theory breaks down. The oscillatory behaviour 
is exactly what one would anticipate from the alternating-sign factorial growth of
the contribution of the leading ${UV}_{1}$ renormalon, given by Eq.(26). {\hspace{.04cm}To} {\hspace{.30cm}attempt}
\newpage
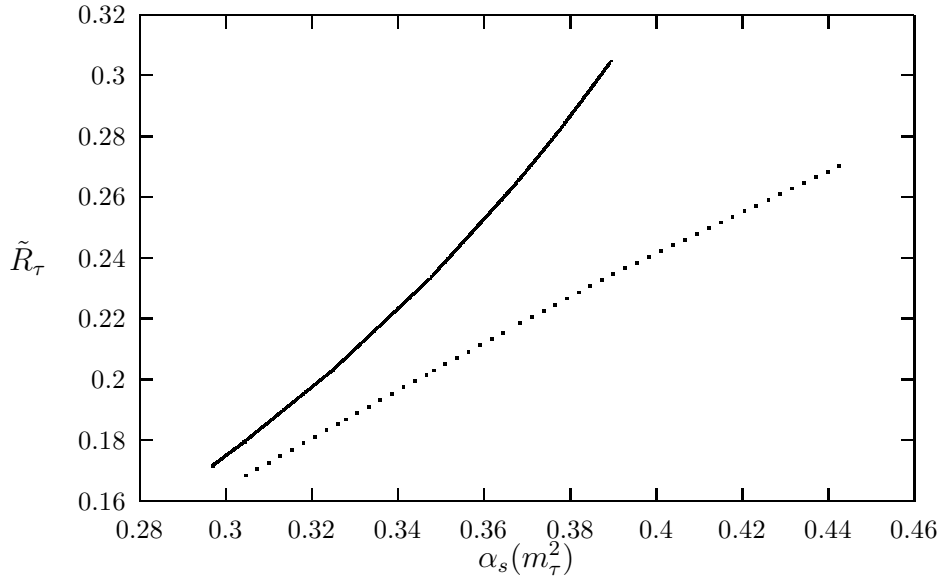
\begin{figure}
\begin{center}
\setlength{\unitlength}{0.240900pt}
\ifx\plotpoint\undefined\newsavebox{\plotpoint}\fi
\sbox{\plotpoint}{\rule[-0.200pt]{0.400pt}{0.400pt}}%
\begin{picture}(1500,900)(0,0)
\font\gnuplot=cmr10 at 10pt
\gnuplot
\sbox{\plotpoint}{\rule[-0.200pt]{0.400pt}{0.400pt}}%
\put(220.0,113.0){\rule[-0.200pt]{4.818pt}{0.400pt}}
\put(198,113){\makebox(0,0)[r]{0.16}}
\put(1416.0,113.0){\rule[-0.200pt]{4.818pt}{0.400pt}}
\put(220.0,209.0){\rule[-0.200pt]{4.818pt}{0.400pt}}
\put(198,209){\makebox(0,0)[r]{0.18}}
\put(1416.0,209.0){\rule[-0.200pt]{4.818pt}{0.400pt}}
\put(220.0,304.0){\rule[-0.200pt]{4.818pt}{0.400pt}}
\put(198,304){\makebox(0,0)[r]{0.2}}
\put(1416.0,304.0){\rule[-0.200pt]{4.818pt}{0.400pt}}
\put(220.0,400.0){\rule[-0.200pt]{4.818pt}{0.400pt}}
\put(198,400){\makebox(0,0)[r]{0.22}}
\put(1416.0,400.0){\rule[-0.200pt]{4.818pt}{0.400pt}}
\put(220.0,495.0){\rule[-0.200pt]{4.818pt}{0.400pt}}
\put(198,495){\makebox(0,0)[r]{0.24}}
\put(1416.0,495.0){\rule[-0.200pt]{4.818pt}{0.400pt}}
\put(220.0,591.0){\rule[-0.200pt]{4.818pt}{0.400pt}}
\put(198,591){\makebox(0,0)[r]{0.26}}
\put(1416.0,591.0){\rule[-0.200pt]{4.818pt}{0.400pt}}
\put(220.0,686.0){\rule[-0.200pt]{4.818pt}{0.400pt}}
\put(198,686){\makebox(0,0)[r]{0.28}}
\put(1416.0,686.0){\rule[-0.200pt]{4.818pt}{0.400pt}}
\put(220.0,782.0){\rule[-0.200pt]{4.818pt}{0.400pt}}
\put(198,782){\makebox(0,0)[r]{0.3}}
\put(1416.0,782.0){\rule[-0.200pt]{4.818pt}{0.400pt}}
\put(220.0,877.0){\rule[-0.200pt]{4.818pt}{0.400pt}}
\put(198,877){\makebox(0,0)[r]{0.32}}
\put(1416.0,877.0){\rule[-0.200pt]{4.818pt}{0.400pt}}
\put(220.0,113.0){\rule[-0.200pt]{0.400pt}{4.818pt}}
\put(220,68){\makebox(0,0){0.28}}
\put(220.0,857.0){\rule[-0.200pt]{0.400pt}{4.818pt}}
\put(355.0,113.0){\rule[-0.200pt]{0.400pt}{4.818pt}}
\put(355,68){\makebox(0,0){0.3}}
\put(355.0,857.0){\rule[-0.200pt]{0.400pt}{4.818pt}}
\put(490.0,113.0){\rule[-0.200pt]{0.400pt}{4.818pt}}
\put(490,68){\makebox(0,0){0.32}}
\put(490.0,857.0){\rule[-0.200pt]{0.400pt}{4.818pt}}
\put(625.0,113.0){\rule[-0.200pt]{0.400pt}{4.818pt}}
\put(625,68){\makebox(0,0){0.34}}
\put(625.0,857.0){\rule[-0.200pt]{0.400pt}{4.818pt}}
\put(760.0,113.0){\rule[-0.200pt]{0.400pt}{4.818pt}}
\put(760,68){\makebox(0,0){0.36}}
\put(760.0,857.0){\rule[-0.200pt]{0.400pt}{4.818pt}}
\put(896.0,113.0){\rule[-0.200pt]{0.400pt}{4.818pt}}
\put(896,68){\makebox(0,0){0.38}}
\put(896.0,857.0){\rule[-0.200pt]{0.400pt}{4.818pt}}
\put(1031.0,113.0){\rule[-0.200pt]{0.400pt}{4.818pt}}
\put(1031,68){\makebox(0,0){0.4}}
\put(1031.0,857.0){\rule[-0.200pt]{0.400pt}{4.818pt}}
\put(1166.0,113.0){\rule[-0.200pt]{0.400pt}{4.818pt}}
\put(1166,68){\makebox(0,0){0.42}}
\put(1166.0,857.0){\rule[-0.200pt]{0.400pt}{4.818pt}}
\put(1301.0,113.0){\rule[-0.200pt]{0.400pt}{4.818pt}}
\put(1301,68){\makebox(0,0){0.44}}
\put(1301.0,857.0){\rule[-0.200pt]{0.400pt}{4.818pt}}
\put(1436.0,113.0){\rule[-0.200pt]{0.400pt}{4.818pt}}
\put(1436,68){\makebox(0,0){0.46}}
\put(1436.0,857.0){\rule[-0.200pt]{0.400pt}{4.818pt}}
\put(220.0,113.0){\rule[-0.200pt]{292.934pt}{0.400pt}}
\put(1436.0,113.0){\rule[-0.200pt]{0.400pt}{184.048pt}}
\put(220.0,877.0){\rule[-0.200pt]{292.934pt}{0.400pt}}
\put(45,495){\makebox(0,0){$\tilde{R}_{\tau}$}}
\put(828,23){\makebox(0,0){${{\alpha}_s}({m^2_{\tau}})$}}
\put(220.0,113.0){\rule[-0.200pt]{0.400pt}{184.048pt}}
\sbox{\plotpoint}{\rule[-0.500pt]{1.000pt}{1.000pt}}%
\put(385,153){\usebox{\plotpoint}}
\multiput(385,153)(18.105,10.148){9}{\usebox{\plotpoint}}
\multiput(542,241)(18.125,10.113){8}{\usebox{\plotpoint}}
\multiput(680,318)(18.231,9.920){7}{\usebox{\plotpoint}}
\multiput(816,392)(18.307,9.780){8}{\usebox{\plotpoint}}
\multiput(962,470)(18.731,8.942){19}{\usebox{\plotpoint}}
\put(1316,639){\usebox{\plotpoint}}
\sbox{\plotpoint}{\rule[-0.400pt]{0.800pt}{0.800pt}}%
\put(334,168){\usebox{\plotpoint}}
\multiput(334.00,169.41)(0.672,0.503){69}{\rule{1.274pt}{0.121pt}}
\multiput(334.00,166.34)(48.356,38.000){2}{\rule{0.637pt}{0.800pt}}
\multiput(385.00,207.41)(0.615,0.501){219}{\rule{1.184pt}{0.121pt}}
\multiput(385.00,204.34)(136.542,113.000){2}{\rule{0.592pt}{0.800pt}}
\multiput(524.00,320.41)(0.530,0.501){285}{\rule{1.049pt}{0.121pt}}
\multiput(524.00,317.34)(152.822,146.000){2}{\rule{0.525pt}{0.800pt}}
\multiput(680.41,465.00)(0.501,0.554){267}{\rule{0.121pt}{1.088pt}}
\multiput(677.34,465.00)(137.000,149.743){2}{\rule{0.800pt}{0.544pt}}
\multiput(817.41,617.00)(0.501,0.618){137}{\rule{0.121pt}{1.189pt}}
\multiput(814.34,617.00)(72.000,86.532){2}{\rule{0.800pt}{0.594pt}}
\multiput(889.41,706.00)(0.501,0.676){141}{\rule{0.121pt}{1.281pt}}
\multiput(886.34,706.00)(74.000,97.341){2}{\rule{0.800pt}{0.641pt}}
\end{picture}
\caption{
${{\tilde{R}}_{\tau}}$ versus ${{\alpha}_{s}({m^2_{\tau}})}$. The dotted curve is
the exact NNLO CORGI fixed-order result, and the upper solid curve is the
approximate all-orders CORGI resummation.
}
\label{fig:Figure2}
\end{center}
\end{figure}
{\hspace{-.7cm}to} estimate the uncertainty in ${\alpha}_{s}({m}_{\tau}^{2})$ extracted from ${R}_{\tau}$
measurements we
can use the difference between the resummed and exact NNLO fixed-order
CORGI results to estimate the possible effects of uncalculated higher order terms.
In Figure 2 we have plotted ${\tilde{R}}_{\tau}$ versus ${\alpha}_{s}({m^2_{\tau}})$.
The upper solid curve is the all-orders CORGI result, whilst the lower dashed curve
is the NNLO fixed-order CORGI result,. We note that the separation of the curves
increases rapidly with increasing ${\tilde{R}}_{\tau}$, so we are fortunate that
for the experimentally measured ${\tilde{R}}_{\tau}{\simeq}0.2$ the  separation of
the curves is reasonably small. Using the ALEPH data we find
${\alpha}_{s}({m^2_{\tau}})
={0.330}^{+0.014}_{-0.013}$
from the all-orders CORGI result, and ${\alpha}_{s}({m^2_{\tau}})
={0.355}^{+0.022}_{-0.022}$ from NNLO fixed order CORGI.
The corresponding results  which would have been obtained by integrating up
the Effective Charge (EC) beta-function for ${\tilde{D}}$ as in Ref.\cite{r19} are
${\alpha}_{s}({m^2_{\tau}})={0.337}^{+0.015}_{-0.016}$ and ${\alpha}_{s}({m^2_{\tau}})=
{0.347}^{+0.021}_{-0.022}$ for the 
resummed and NNLO EC results.
So, as expected, the two approaches yield similar results.
If we evolve these ${\alpha}_{s}({m}_{\tau}^{2})$
results through flavour thresholds up to ${\mu}={M_Z}$ {\hspace{.3cm}using} 
\newpage
\begin{figure}
\begin{center}
\setlength{\unitlength}{0.240900pt}
\ifx\plotpoint\undefined\newsavebox{\plotpoint}\fi
\sbox{\plotpoint}{\rule[-0.200pt]{0.400pt}{0.400pt}}%
\begin{picture}(1500,900)(0,0)
\font\gnuplot=cmr10 at 10pt
\gnuplot
\sbox{\plotpoint}{\rule[-0.200pt]{0.400pt}{0.400pt}}%
\put(220.0,113.0){\rule[-0.200pt]{4.818pt}{0.400pt}}
\put(198,113){\makebox(0,0)[r]{0.16}}
\put(1416.0,113.0){\rule[-0.200pt]{4.818pt}{0.400pt}}
\put(220.0,209.0){\rule[-0.200pt]{4.818pt}{0.400pt}}
\put(198,209){\makebox(0,0)[r]{0.18}}
\put(1416.0,209.0){\rule[-0.200pt]{4.818pt}{0.400pt}}
\put(220.0,304.0){\rule[-0.200pt]{4.818pt}{0.400pt}}
\put(198,304){\makebox(0,0)[r]{0.2}}
\put(1416.0,304.0){\rule[-0.200pt]{4.818pt}{0.400pt}}
\put(220.0,400.0){\rule[-0.200pt]{4.818pt}{0.400pt}}
\put(198,400){\makebox(0,0)[r]{0.22}}
\put(1416.0,400.0){\rule[-0.200pt]{4.818pt}{0.400pt}}
\put(220.0,495.0){\rule[-0.200pt]{4.818pt}{0.400pt}}
\put(198,495){\makebox(0,0)[r]{0.24}}
\put(1416.0,495.0){\rule[-0.200pt]{4.818pt}{0.400pt}}
\put(220.0,591.0){\rule[-0.200pt]{4.818pt}{0.400pt}}
\put(198,591){\makebox(0,0)[r]{0.26}}
\put(1416.0,591.0){\rule[-0.200pt]{4.818pt}{0.400pt}}
\put(220.0,686.0){\rule[-0.200pt]{4.818pt}{0.400pt}}
\put(198,686){\makebox(0,0)[r]{0.28}}
\put(1416.0,686.0){\rule[-0.200pt]{4.818pt}{0.400pt}}
\put(220.0,782.0){\rule[-0.200pt]{4.818pt}{0.400pt}}
\put(198,782){\makebox(0,0)[r]{0.3}}
\put(1416.0,782.0){\rule[-0.200pt]{4.818pt}{0.400pt}}
\put(220.0,877.0){\rule[-0.200pt]{4.818pt}{0.400pt}}
\put(198,877){\makebox(0,0)[r]{0.32}}
\put(1416.0,877.0){\rule[-0.200pt]{4.818pt}{0.400pt}}
\put(220.0,113.0){\rule[-0.200pt]{0.400pt}{4.818pt}}
\put(220,68){\makebox(0,0){0.114}}
\put(220.0,857.0){\rule[-0.200pt]{0.400pt}{4.818pt}}
\put(355.0,113.0){\rule[-0.200pt]{0.400pt}{4.818pt}}
\put(355,68){\makebox(0,0){0.116}}
\put(355.0,857.0){\rule[-0.200pt]{0.400pt}{4.818pt}}
\put(490.0,113.0){\rule[-0.200pt]{0.400pt}{4.818pt}}
\put(490,68){\makebox(0,0){0.118}}
\put(490.0,857.0){\rule[-0.200pt]{0.400pt}{4.818pt}}
\put(625.0,113.0){\rule[-0.200pt]{0.400pt}{4.818pt}}
\put(625,68){\makebox(0,0){0.12}}
\put(625.0,857.0){\rule[-0.200pt]{0.400pt}{4.818pt}}
\put(760.0,113.0){\rule[-0.200pt]{0.400pt}{4.818pt}}
\put(760,68){\makebox(0,0){0.122}}
\put(760.0,857.0){\rule[-0.200pt]{0.400pt}{4.818pt}}
\put(896.0,113.0){\rule[-0.200pt]{0.400pt}{4.818pt}}
\put(896,68){\makebox(0,0){0.124}}
\put(896.0,857.0){\rule[-0.200pt]{0.400pt}{4.818pt}}
\put(1031.0,113.0){\rule[-0.200pt]{0.400pt}{4.818pt}}
\put(1031,68){\makebox(0,0){0.126}}
\put(1031.0,857.0){\rule[-0.200pt]{0.400pt}{4.818pt}}
\put(1166.0,113.0){\rule[-0.200pt]{0.400pt}{4.818pt}}
\put(1166,68){\makebox(0,0){0.128}}
\put(1166.0,857.0){\rule[-0.200pt]{0.400pt}{4.818pt}}
\put(1301.0,113.0){\rule[-0.200pt]{0.400pt}{4.818pt}}
\put(1301,68){\makebox(0,0){0.13}}
\put(1301.0,857.0){\rule[-0.200pt]{0.400pt}{4.818pt}}
\put(1436.0,113.0){\rule[-0.200pt]{0.400pt}{4.818pt}}
\put(1436,68){\makebox(0,0){0.132}}
\put(1436.0,857.0){\rule[-0.200pt]{0.400pt}{4.818pt}}
\put(220.0,113.0){\rule[-0.200pt]{292.934pt}{0.400pt}}
\put(1436.0,113.0){\rule[-0.200pt]{0.400pt}{184.048pt}}
\put(220.0,877.0){\rule[-0.200pt]{292.934pt}{0.400pt}}
\put(45,495){\makebox(0,0){$\tilde{R}_{\tau}$}}
\put(828,23){\makebox(0,0){${{\alpha}_s}({M^2_{Z}})$}}
\put(220.0,113.0){\rule[-0.200pt]{0.400pt}{184.048pt}}
\sbox{\plotpoint}{\rule[-0.400pt]{0.800pt}{0.800pt}}%
\put(335,168){\usebox{\plotpoint}}
\multiput(335.00,169.41)(0.927,0.503){69}{\rule{1.674pt}{0.121pt}}
\multiput(335.00,166.34)(66.526,38.000){2}{\rule{0.837pt}{0.800pt}}
\multiput(405.00,207.41)(0.780,0.501){219}{\rule{1.446pt}{0.121pt}}
\multiput(405.00,204.34)(172.999,113.000){2}{\rule{0.723pt}{0.800pt}}
\multiput(581.00,320.41)(0.606,0.501){285}{\rule{1.170pt}{0.121pt}}
\multiput(581.00,317.34)(174.572,146.000){2}{\rule{0.585pt}{0.800pt}}
\multiput(759.41,465.00)(0.501,0.531){279}{\rule{0.121pt}{1.050pt}}
\multiput(756.34,465.00)(143.000,149.820){2}{\rule{0.800pt}{0.525pt}}
\multiput(902.41,617.00)(0.501,0.645){131}{\rule{0.121pt}{1.232pt}}
\multiput(899.34,617.00)(69.000,86.443){2}{\rule{0.800pt}{0.616pt}}
\multiput(971.41,706.00)(0.501,0.726){131}{\rule{0.121pt}{1.359pt}}
\multiput(968.34,706.00)(69.000,97.178){2}{\rule{0.800pt}{0.680pt}}
\sbox{\plotpoint}{\rule[-0.500pt]{1.000pt}{1.000pt}}%
\put(405,153){\usebox{\plotpoint}}
\multiput(405,153)(18.951,8.465){11}{\usebox{\plotpoint}}
\multiput(602,241)(18.658,9.093){8}{\usebox{\plotpoint}}
\multiput(760,318)(18.378,9.645){8}{\usebox{\plotpoint}}
\multiput(901,392)(18.069,10.213){8}{\usebox{\plotpoint}}
\multiput(1039,470)(17.836,10.614){16}{\usebox{\plotpoint}}
\put(1323,639){\usebox{\plotpoint}}
\end{picture}
\caption{
As Figure 2 but versus ${\alpha}_{s}({M}_{Z}^{2})$.
}
\label{fig:Figure3}
\end{center}
\end{figure}
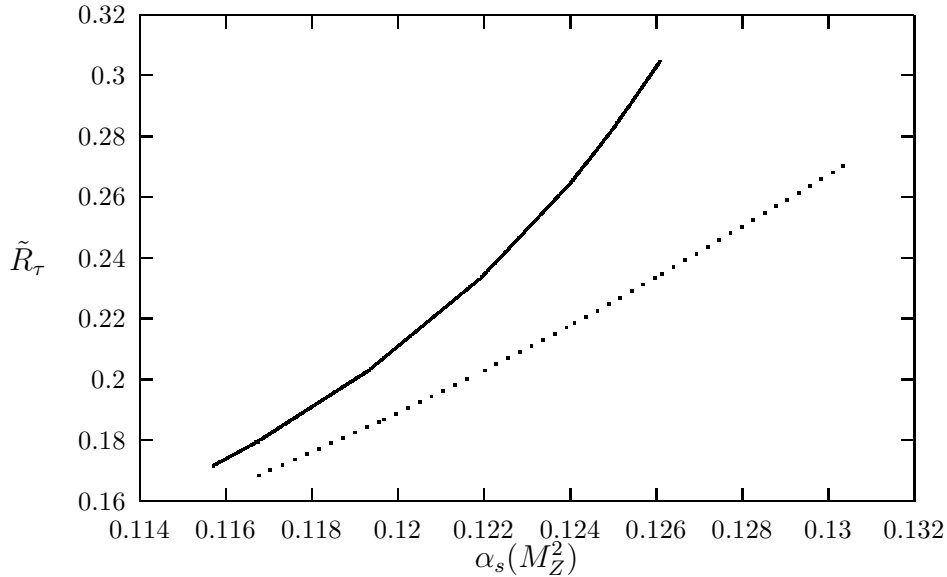
{\hspace{-.7cm}the} three-loop matching
conditions \cite{r32,r33}, we find {\hspace{.15cm}${{\alpha}_{s}}({M^2_Z})$=${0.120}^{+0.002}_{-0.022}$} from the resummed
CORGI result, and ${{\alpha}_{s}}({M^2_{Z}})={0.123}^{+0.002}_{-0.002}$ from the NNLO CORGI
result. Thus, we conservatively estimate an uncertainty ${\delta}{\alpha}_{s}(M^2_{Z}){\approx}0.003$. A
direct plot of the resummed and NNLO results for ${\tilde{R}}_{\tau}$ versus ${\alpha}_{s}(M^2_{Z})$
is given in Figure 3.
The invariant mass distribution of the produced hadrons in ${\tau}$ decay is
well-measured experimentally \cite{r22,r23}. 
\newline
We define the quantity ${R}_{\tau}({s}_{0})$ as
\be
{R}_{\tau}({s_0}){\equiv}\frac{{\Gamma}({\tau}{\rightarrow}{\nu}_{\tau}+{\rm{hadrons}};{s}_{had}>{s_0})}
{{\Gamma}({\tau}{\rightarrow}{\nu}_{\tau}{e}{\bar{\nu}}_{e})}={\int_{0}^{s_0}}{ds}\frac{d{R}_{\tau}(s)}
{ds}\;,
\ee
where $\frac{d{R}_{\tau}}{ds}$ denotes the measured inclusive hadronic spectrum. 
\be
{R}_{\tau}({s_0})=N({{\vert}{V_{ud}}\vert}^2){S_{EW}}[(2x-2x^3+x^4)
+{\frac{3}{4}}{c_F}{{\tilde{R}}_{\tau}({s}_{0})}+{\delta}_{PC}]\;,
\ee
with $x{\equiv}{s_0}/{m^2_{\tau}}$. The perturbative part ${\tilde{R}}_{\tau}({s}_{0})$
can be computed from Eq.(10) with the choice of weight function
\be
W({\theta})=2x(1+{e}^{i{\theta}})-2{x}^{3}(1+{e}^{3i{\theta}})+{x}^{4}(1-{e}^{4i{\theta}})\;. 
\ee
\newpage
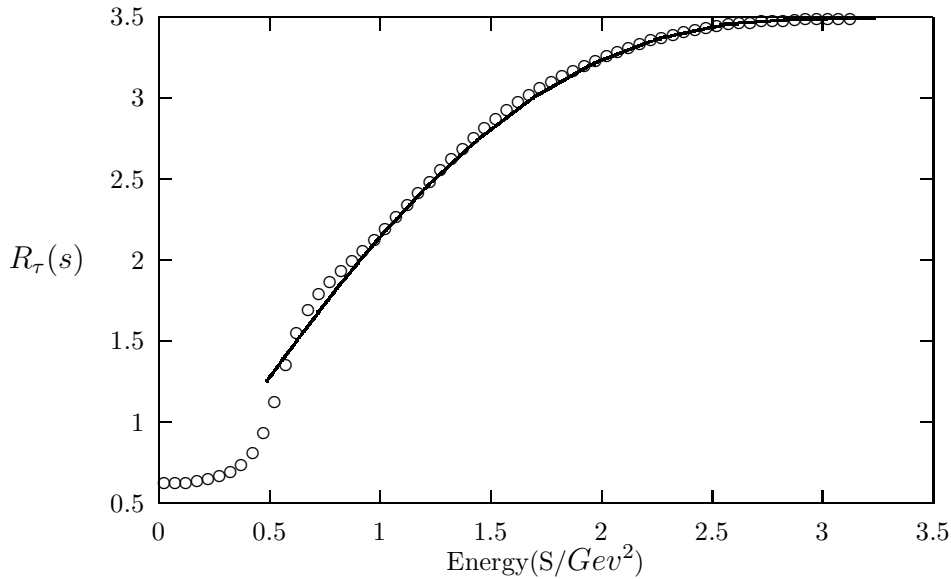
\begin{figure}
\begin{center}
\setlength{\unitlength}{0.240900pt}
\ifx\plotpoint\undefined\newsavebox{\plotpoint}\fi
\sbox{\plotpoint}{\rule[-0.200pt]{0.400pt}{0.400pt}}%
\begin{picture}(1500,900)(0,0)
\font\gnuplot=cmr10 at 10pt
\gnuplot
\sbox{\plotpoint}{\rule[-0.200pt]{0.400pt}{0.400pt}}%
\put(220.0,113.0){\rule[-0.200pt]{0.400pt}{184.048pt}}
\put(220.0,113.0){\rule[-0.200pt]{4.818pt}{0.400pt}}
\put(198,113){\makebox(0,0)[r]{0.5}}
\put(1416.0,113.0){\rule[-0.200pt]{4.818pt}{0.400pt}}
\put(220.0,240.0){\rule[-0.200pt]{4.818pt}{0.400pt}}
\put(198,240){\makebox(0,0)[r]{1}}
\put(1416.0,240.0){\rule[-0.200pt]{4.818pt}{0.400pt}}
\put(220.0,368.0){\rule[-0.200pt]{4.818pt}{0.400pt}}
\put(198,368){\makebox(0,0)[r]{1.5}}
\put(1416.0,368.0){\rule[-0.200pt]{4.818pt}{0.400pt}}
\put(220.0,495.0){\rule[-0.200pt]{4.818pt}{0.400pt}}
\put(198,495){\makebox(0,0)[r]{2}}
\put(1416.0,495.0){\rule[-0.200pt]{4.818pt}{0.400pt}}
\put(220.0,622.0){\rule[-0.200pt]{4.818pt}{0.400pt}}
\put(198,622){\makebox(0,0)[r]{2.5}}
\put(1416.0,622.0){\rule[-0.200pt]{4.818pt}{0.400pt}}
\put(220.0,750.0){\rule[-0.200pt]{4.818pt}{0.400pt}}
\put(198,750){\makebox(0,0)[r]{3}}
\put(1416.0,750.0){\rule[-0.200pt]{4.818pt}{0.400pt}}
\put(220.0,877.0){\rule[-0.200pt]{4.818pt}{0.400pt}}
\put(198,877){\makebox(0,0)[r]{3.5}}
\put(1416.0,877.0){\rule[-0.200pt]{4.818pt}{0.400pt}}
\put(220.0,113.0){\rule[-0.200pt]{0.400pt}{4.818pt}}
\put(220,68){\makebox(0,0){0}}
\put(220.0,857.0){\rule[-0.200pt]{0.400pt}{4.818pt}}
\put(394.0,113.0){\rule[-0.200pt]{0.400pt}{4.818pt}}
\put(394,68){\makebox(0,0){0.5}}
\put(394.0,857.0){\rule[-0.200pt]{0.400pt}{4.818pt}}
\put(567.0,113.0){\rule[-0.200pt]{0.400pt}{4.818pt}}
\put(567,68){\makebox(0,0){1}}
\put(567.0,857.0){\rule[-0.200pt]{0.400pt}{4.818pt}}
\put(741.0,113.0){\rule[-0.200pt]{0.400pt}{4.818pt}}
\put(741,68){\makebox(0,0){1.5}}
\put(741.0,857.0){\rule[-0.200pt]{0.400pt}{4.818pt}}
\put(915.0,113.0){\rule[-0.200pt]{0.400pt}{4.818pt}}
\put(915,68){\makebox(0,0){2}}
\put(915.0,857.0){\rule[-0.200pt]{0.400pt}{4.818pt}}
\put(1089.0,113.0){\rule[-0.200pt]{0.400pt}{4.818pt}}
\put(1089,68){\makebox(0,0){2.5}}
\put(1089.0,857.0){\rule[-0.200pt]{0.400pt}{4.818pt}}
\put(1262.0,113.0){\rule[-0.200pt]{0.400pt}{4.818pt}}
\put(1262,68){\makebox(0,0){3}}
\put(1262.0,857.0){\rule[-0.200pt]{0.400pt}{4.818pt}}
\put(1436.0,113.0){\rule[-0.200pt]{0.400pt}{4.818pt}}
\put(1436,68){\makebox(0,0){3.5}}
\put(1436.0,857.0){\rule[-0.200pt]{0.400pt}{4.818pt}}
\put(220.0,113.0){\rule[-0.200pt]{292.934pt}{0.400pt}}
\put(1436.0,113.0){\rule[-0.200pt]{0.400pt}{184.048pt}}
\put(220.0,877.0){\rule[-0.200pt]{292.934pt}{0.400pt}}
\put(45,495){\makebox(0,0){${R_{\tau}(s)}$}}
\put(828,23){\makebox(0,0){Energy(S/${Gev^{2}}$)}}
\put(220.0,113.0){\rule[-0.200pt]{0.400pt}{184.048pt}}
\put(229,146){\raisebox{-.8pt}{\makebox(0,0){$\circ$}}}
\put(246,146){\raisebox{-.8pt}{\makebox(0,0){$\circ$}}}
\put(263,147){\raisebox{-.8pt}{\makebox(0,0){$\circ$}}}
\put(281,149){\raisebox{-.8pt}{\makebox(0,0){$\circ$}}}
\put(298,152){\raisebox{-.8pt}{\makebox(0,0){$\circ$}}}
\put(316,157){\raisebox{-.8pt}{\makebox(0,0){$\circ$}}}
\put(333,164){\raisebox{-.8pt}{\makebox(0,0){$\circ$}}}
\put(350,175){\raisebox{-.8pt}{\makebox(0,0){$\circ$}}}
\put(368,194){\raisebox{-.8pt}{\makebox(0,0){$\circ$}}}
\put(385,225){\raisebox{-.8pt}{\makebox(0,0){$\circ$}}}
\put(402,273){\raisebox{-.8pt}{\makebox(0,0){$\circ$}}}
\put(420,332){\raisebox{-.8pt}{\makebox(0,0){$\circ$}}}
\put(437,382){\raisebox{-.8pt}{\makebox(0,0){$\circ$}}}
\put(455,418){\raisebox{-.8pt}{\makebox(0,0){$\circ$}}}
\put(472,443){\raisebox{-.8pt}{\makebox(0,0){$\circ$}}}
\put(489,462){\raisebox{-.8pt}{\makebox(0,0){$\circ$}}}
\put(507,479){\raisebox{-.8pt}{\makebox(0,0){$\circ$}}}
\put(524,495){\raisebox{-.8pt}{\makebox(0,0){$\circ$}}}
\put(541,511){\raisebox{-.8pt}{\makebox(0,0){$\circ$}}}
\put(559,528){\raisebox{-.8pt}{\makebox(0,0){$\circ$}}}
\put(576,546){\raisebox{-.8pt}{\makebox(0,0){$\circ$}}}
\put(593,564){\raisebox{-.8pt}{\makebox(0,0){$\circ$}}}
\put(611,583){\raisebox{-.8pt}{\makebox(0,0){$\circ$}}}
\put(628,602){\raisebox{-.8pt}{\makebox(0,0){$\circ$}}}
\put(646,620){\raisebox{-.8pt}{\makebox(0,0){$\circ$}}}
\put(663,638){\raisebox{-.8pt}{\makebox(0,0){$\circ$}}}
\put(680,655){\raisebox{-.8pt}{\makebox(0,0){$\circ$}}}
\put(698,672){\raisebox{-.8pt}{\makebox(0,0){$\circ$}}}
\put(715,689){\raisebox{-.8pt}{\makebox(0,0){$\circ$}}}
\put(732,704){\raisebox{-.8pt}{\makebox(0,0){$\circ$}}}
\put(750,719){\raisebox{-.8pt}{\makebox(0,0){$\circ$}}}
\put(767,732){\raisebox{-.8pt}{\makebox(0,0){$\circ$}}}
\put(785,745){\raisebox{-.8pt}{\makebox(0,0){$\circ$}}}
\put(802,756){\raisebox{-.8pt}{\makebox(0,0){$\circ$}}}
\put(819,767){\raisebox{-.8pt}{\makebox(0,0){$\circ$}}}
\put(837,777){\raisebox{-.8pt}{\makebox(0,0){$\circ$}}}
\put(854,786){\raisebox{-.8pt}{\makebox(0,0){$\circ$}}}
\put(871,794){\raisebox{-.8pt}{\makebox(0,0){$\circ$}}}
\put(889,802){\raisebox{-.8pt}{\makebox(0,0){$\circ$}}}
\put(906,810){\raisebox{-.8pt}{\makebox(0,0){$\circ$}}}
\put(924,817){\raisebox{-.8pt}{\makebox(0,0){$\circ$}}}
\put(941,823){\raisebox{-.8pt}{\makebox(0,0){$\circ$}}}
\put(958,830){\raisebox{-.8pt}{\makebox(0,0){$\circ$}}}
\put(976,836){\raisebox{-.8pt}{\makebox(0,0){$\circ$}}}
\put(993,842){\raisebox{-.8pt}{\makebox(0,0){$\circ$}}}
\put(1010,846){\raisebox{-.8pt}{\makebox(0,0){$\circ$}}}
\put(1028,851){\raisebox{-.8pt}{\makebox(0,0){$\circ$}}}
\put(1045,855){\raisebox{-.8pt}{\makebox(0,0){$\circ$}}}
\put(1063,858){\raisebox{-.8pt}{\makebox(0,0){$\circ$}}}
\put(1080,862){\raisebox{-.8pt}{\makebox(0,0){$\circ$}}}
\put(1097,864){\raisebox{-.8pt}{\makebox(0,0){$\circ$}}}
\put(1115,867){\raisebox{-.8pt}{\makebox(0,0){$\circ$}}}
\put(1132,869){\raisebox{-.8pt}{\makebox(0,0){$\circ$}}}
\put(1149,870){\raisebox{-.8pt}{\makebox(0,0){$\circ$}}}
\put(1167,872){\raisebox{-.8pt}{\makebox(0,0){$\circ$}}}
\put(1184,873){\raisebox{-.8pt}{\makebox(0,0){$\circ$}}}
\put(1201,873){\raisebox{-.8pt}{\makebox(0,0){$\circ$}}}
\put(1219,874){\raisebox{-.8pt}{\makebox(0,0){$\circ$}}}
\put(1236,875){\raisebox{-.8pt}{\makebox(0,0){$\circ$}}}
\put(1254,875){\raisebox{-.8pt}{\makebox(0,0){$\circ$}}}
\put(1271,875){\raisebox{-.8pt}{\makebox(0,0){$\circ$}}}
\put(1288,875){\raisebox{-.8pt}{\makebox(0,0){$\circ$}}}
\put(1306,875){\raisebox{-.8pt}{\makebox(0,0){$\circ$}}}
\sbox{\plotpoint}{\rule[-0.400pt]{0.800pt}{0.800pt}}%
\put(390,304){\usebox{\plotpoint}}
\multiput(391.41,304.00)(0.504,0.682){43}{\rule{0.121pt}{1.288pt}}
\multiput(388.34,304.00)(25.000,31.327){2}{\rule{0.800pt}{0.644pt}}
\multiput(416.41,338.00)(0.504,0.668){47}{\rule{0.121pt}{1.267pt}}
\multiput(413.34,338.00)(27.000,33.371){2}{\rule{0.800pt}{0.633pt}}
\multiput(443.41,374.00)(0.504,0.656){51}{\rule{0.121pt}{1.248pt}}
\multiput(440.34,374.00)(29.000,35.409){2}{\rule{0.800pt}{0.624pt}}
\multiput(472.41,412.00)(0.503,0.651){53}{\rule{0.121pt}{1.240pt}}
\multiput(469.34,412.00)(30.000,36.426){2}{\rule{0.800pt}{0.620pt}}
\multiput(502.41,451.00)(0.503,0.606){59}{\rule{0.121pt}{1.170pt}}
\multiput(499.34,451.00)(33.000,37.572){2}{\rule{0.800pt}{0.585pt}}
\multiput(535.41,491.00)(0.503,0.590){59}{\rule{0.121pt}{1.145pt}}
\multiput(532.34,491.00)(33.000,36.623){2}{\rule{0.800pt}{0.573pt}}
\multiput(568.41,530.00)(0.501,0.540){139}{\rule{0.121pt}{1.066pt}}
\multiput(565.34,530.00)(73.000,76.788){2}{\rule{0.800pt}{0.533pt}}
\multiput(640.00,610.41)(0.540,0.501){141}{\rule{1.065pt}{0.121pt}}
\multiput(640.00,607.34)(77.790,74.000){2}{\rule{0.532pt}{0.800pt}}
\multiput(720.00,684.41)(0.660,0.501){125}{\rule{1.255pt}{0.121pt}}
\multiput(720.00,681.34)(84.396,66.000){2}{\rule{0.627pt}{0.800pt}}
\multiput(807.00,750.41)(0.874,0.502){101}{\rule{1.593pt}{0.121pt}}
\multiput(807.00,747.34)(90.694,54.000){2}{\rule{0.796pt}{0.800pt}}
\multiput(901.00,804.41)(1.308,0.503){71}{\rule{2.272pt}{0.121pt}}
\multiput(901.00,801.34)(96.285,39.000){2}{\rule{1.136pt}{0.800pt}}
\multiput(1002.00,843.41)(2.387,0.505){39}{\rule{3.922pt}{0.122pt}}
\multiput(1002.00,840.34)(98.860,23.000){2}{\rule{1.961pt}{0.800pt}}
\multiput(1109.00,866.40)(7.174,0.516){11}{\rule{10.422pt}{0.124pt}}
\multiput(1109.00,863.34)(93.368,9.000){2}{\rule{5.211pt}{0.800pt}}
\put(1224,872.84){\rule{22.404pt}{0.800pt}}
\multiput(1224.00,872.34)(46.500,1.000){2}{\rule{11.202pt}{0.800pt}}
\put(1317.0,875.0){\rule[-0.400pt]{6.986pt}{0.800pt}}
\end{picture}
\caption{
ALEPH data for ${R}_{\tau}(s)$ (open circles) compared with leading-$b$
all-orders CORGI result fitted at $s={m}_{\tau}^{2}$ (solid curve).
}
\label{fig:Figure4}
\end{center}
\end{figure}
{\hspace{-.7cm}It} is then straightforward to obtain contour-improved fixed-order and resummed
CORGI results for ${\tilde{R}}({s_0})$. In Figure 4 we show the fit of
the all-orders leading-$b$ CORGI resummation (solid line) to the ALEPH data
for ${R}_{\tau}(s)$ (open circles) \cite{r22}. The resummation is fitted to the data
at ${s}={m}_{\tau}^{2}$, where ${R}_{\tau}({m}_{\tau}^{2})={R}_{\tau}$. The
CORGI coupling has a Landau pole at ${\sqrt{s}}={\Lambda}_{D}$, as is apparent
from Eq.(17). Fitting to the experimental value of ${R}_{\tau}$ determines
${\Lambda}_{D}=0.725$ GeV, and so the resummed prediction is only defined for
$s>0.525\;{{\rm{GeV}}^{2}}$. There is excellent agreement with the data. On this
scale the fixed-order NNLO CORGI result would not be distinguishable from the
all-orders result, and so we have not included it on the plot.
\section*{5 Estimating the uncertainty in hadronic corrections to
${\alpha}({M^2_{Z}})$}
\newpage
\begin{figure}
\begin{center}
\setlength{\unitlength}{0.240900pt}
\ifx\plotpoint\undefined\newsavebox{\plotpoint}\fi
\sbox{\plotpoint}{\rule[-0.200pt]{0.400pt}{0.400pt}}%
\begin{picture}(1500,900)(0,0)
\font\gnuplot=cmr10 at 10pt
\gnuplot
\sbox{\plotpoint}{\rule[-0.200pt]{0.400pt}{0.400pt}}%
\put(220.0,113.0){\rule[-0.200pt]{4.818pt}{0.400pt}}
\put(198,113){\makebox(0,0)[r]{0.08}}
\put(1416.0,113.0){\rule[-0.200pt]{4.818pt}{0.400pt}}
\put(220.0,287.0){\rule[-0.200pt]{4.818pt}{0.400pt}}
\put(198,287){\makebox(0,0)[r]{0.085}}
\put(1416.0,287.0){\rule[-0.200pt]{4.818pt}{0.400pt}}
\put(220.0,460.0){\rule[-0.200pt]{4.818pt}{0.400pt}}
\put(198,460){\makebox(0,0)[r]{0.09}}
\put(1416.0,460.0){\rule[-0.200pt]{4.818pt}{0.400pt}}
\put(220.0,634.0){\rule[-0.200pt]{4.818pt}{0.400pt}}
\put(198,634){\makebox(0,0)[r]{0.095}}
\put(1416.0,634.0){\rule[-0.200pt]{4.818pt}{0.400pt}}
\put(220.0,808.0){\rule[-0.200pt]{4.818pt}{0.400pt}}
\put(198,808){\makebox(0,0)[r]{0.1}}
\put(1416.0,808.0){\rule[-0.200pt]{4.818pt}{0.400pt}}
\put(220.0,113.0){\rule[-0.200pt]{0.400pt}{4.818pt}}
\put(220,68){\makebox(0,0){0}}
\put(220.0,857.0){\rule[-0.200pt]{0.400pt}{4.818pt}}
\put(407.0,113.0){\rule[-0.200pt]{0.400pt}{4.818pt}}
\put(407,68){\makebox(0,0){2}}
\put(407.0,857.0){\rule[-0.200pt]{0.400pt}{4.818pt}}
\put(594.0,113.0){\rule[-0.200pt]{0.400pt}{4.818pt}}
\put(594,68){\makebox(0,0){4}}
\put(594.0,857.0){\rule[-0.200pt]{0.400pt}{4.818pt}}
\put(781.0,113.0){\rule[-0.200pt]{0.400pt}{4.818pt}}
\put(781,68){\makebox(0,0){6}}
\put(781.0,857.0){\rule[-0.200pt]{0.400pt}{4.818pt}}
\put(968.0,113.0){\rule[-0.200pt]{0.400pt}{4.818pt}}
\put(968,68){\makebox(0,0){8}}
\put(968.0,857.0){\rule[-0.200pt]{0.400pt}{4.818pt}}
\put(1155.0,113.0){\rule[-0.200pt]{0.400pt}{4.818pt}}
\put(1155,68){\makebox(0,0){10}}
\put(1155.0,857.0){\rule[-0.200pt]{0.400pt}{4.818pt}}
\put(1342.0,113.0){\rule[-0.200pt]{0.400pt}{4.818pt}}
\put(1342,68){\makebox(0,0){12}}
\put(1342.0,857.0){\rule[-0.200pt]{0.400pt}{4.818pt}}
\put(220.0,113.0){\rule[-0.200pt]{292.934pt}{0.400pt}}
\put(1436.0,113.0){\rule[-0.200pt]{0.400pt}{184.048pt}}
\put(220.0,877.0){\rule[-0.200pt]{292.934pt}{0.400pt}}
\put(45,495){\makebox(0,0){${\tilde{R}}$}}
\put(828,23){\makebox(0,0){Order of pert.theory}}
\put(220.0,113.0){\rule[-0.200pt]{0.400pt}{184.048pt}}
\put(407,701){\raisebox{-.8pt}{\makebox(0,0){$\star$}}}
\put(501,672){\raisebox{-.8pt}{\makebox(0,0){$\star$}}}
\put(594,680){\raisebox{-.8pt}{\makebox(0,0){$\star$}}}
\put(688,696){\raisebox{-.8pt}{\makebox(0,0){$\star$}}}
\put(781,660){\raisebox{-.8pt}{\makebox(0,0){$\star$}}}
\put(875,697){\raisebox{-.8pt}{\makebox(0,0){$\star$}}}
\put(968,679){\raisebox{-.8pt}{\makebox(0,0){$\star$}}}
\put(1062,632){\raisebox{-.8pt}{\makebox(0,0){$\star$}}}
\put(1155,812){\raisebox{-.8pt}{\makebox(0,0){$\star$}}}
\put(1249,463){\raisebox{-.8pt}{\makebox(0,0){$\star$}}}
\put(1342,783){\raisebox{-.8pt}{\makebox(0,0){$\star$}}}
\put(220,651){\usebox{\plotpoint}}
\put(220.0,651.0){\rule[-0.200pt]{292.934pt}{0.400pt}}
\end{picture}
\caption{Fixed order results (starred points) 
for ${
\tilde{R}}$ versus different orders of perturbation theory
at ${\sqrt{s}}={M_{\tau}}$=1.777 GeV. The solid line shows ${\tilde{R}}$ for 
the all-orders contour-improved resummation.
}
\label{fig:Figure5}
\end{center}
\end{figure}
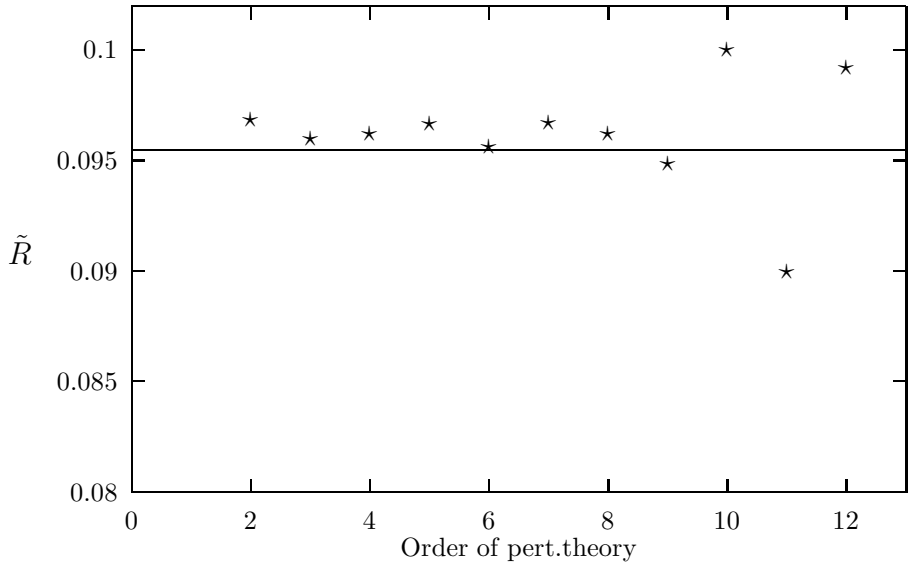
In this section we wish to make use of the difference between the NNLO fixed-order
and resummed CORGI results for ${\tilde{R}}(s)$ in ${e}^{+}{e}^{-}$ annihilation to
estimate the uncertainty in ${\alpha}({M^2_Z})$, the QED coupling at the $Z$ pole,
which plays a crucial role in constraining the Standard Model Higgs mass from
precision electroweak fits to radiative corrections \cite{r34}. We begin, however, by plotting
some figures, analogous to Figure 1, to indicate the performance of fixed-order
perturbation theory versus the resummed results at various energies. In Figure 5
we show the all-orders CORGI leading-$b$ resummation (solid line) and fixed order
results (starred points) for ${\tilde{R}}(s)$ at
${\sqrt{s}}=1.777$ GeV, corresponding to ${m}_{\tau}$,
so the performance can be directly compared to Figure 1. The only difference in the
two calculations is the choice of weight function, $W(\theta)$ in Eq.(10). The oscillatory trend due
to the leading ultraviolet renormalon is again evident, with wild oscillations
setting in at $n>9$ where fixed-order perturbation theory breaks down.\\
In Figure 6 we present a corresponding plot {\hspace{0.2cm}at} {\hspace{0.2cm}LEP1} {\hspace{.2cm}energy} 
{\hspace{.20cm}${\sqrt{s}}={M_Z}$}.
\newpage
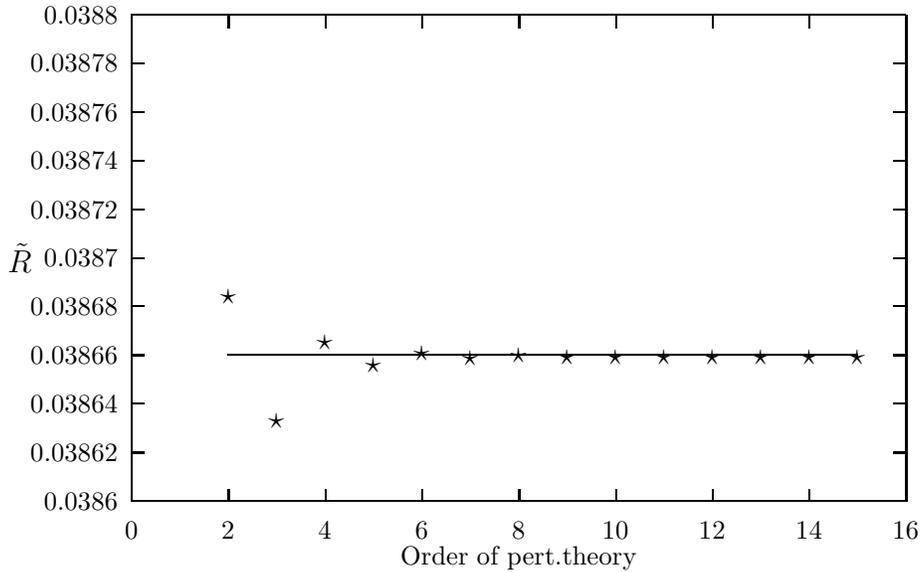
\begin{figure}
\begin{center}
\setlength{\unitlength}{0.240900pt}
\ifx\plotpoint\undefined\newsavebox{\plotpoint}\fi
\sbox{\plotpoint}{\rule[-0.200pt]{0.400pt}{0.400pt}}%
\begin{picture}(1500,900)(0,0)
\font\gnuplot=cmr10 at 10pt
\gnuplot
\sbox{\plotpoint}{\rule[-0.200pt]{0.400pt}{0.400pt}}%
\put(220.0,113.0){\rule[-0.200pt]{4.818pt}{0.400pt}}
\put(198,113){\makebox(0,0)[r]{0.0386}}
\put(1416.0,113.0){\rule[-0.200pt]{4.818pt}{0.400pt}}
\put(220.0,189.0){\rule[-0.200pt]{4.818pt}{0.400pt}}
\put(198,189){\makebox(0,0)[r]{0.03862}}
\put(1416.0,189.0){\rule[-0.200pt]{4.818pt}{0.400pt}}
\put(220.0,266.0){\rule[-0.200pt]{4.818pt}{0.400pt}}
\put(198,266){\makebox(0,0)[r]{0.03864}}
\put(1416.0,266.0){\rule[-0.200pt]{4.818pt}{0.400pt}}
\put(220.0,342.0){\rule[-0.200pt]{4.818pt}{0.400pt}}
\put(198,342){\makebox(0,0)[r]{0.03866}}
\put(1416.0,342.0){\rule[-0.200pt]{4.818pt}{0.400pt}}
\put(220.0,419.0){\rule[-0.200pt]{4.818pt}{0.400pt}}
\put(198,419){\makebox(0,0)[r]{0.03868}}
\put(1416.0,419.0){\rule[-0.200pt]{4.818pt}{0.400pt}}
\put(220.0,495.0){\rule[-0.200pt]{4.818pt}{0.400pt}}
\put(198,495){\makebox(0,0)[r]{0.0387}}
\put(1416.0,495.0){\rule[-0.200pt]{4.818pt}{0.400pt}}
\put(220.0,571.0){\rule[-0.200pt]{4.818pt}{0.400pt}}
\put(198,571){\makebox(0,0)[r]{0.03872}}
\put(1416.0,571.0){\rule[-0.200pt]{4.818pt}{0.400pt}}
\put(220.0,648.0){\rule[-0.200pt]{4.818pt}{0.400pt}}
\put(198,648){\makebox(0,0)[r]{0.03874}}
\put(1416.0,648.0){\rule[-0.200pt]{4.818pt}{0.400pt}}
\put(220.0,724.0){\rule[-0.200pt]{4.818pt}{0.400pt}}
\put(198,724){\makebox(0,0)[r]{0.03876}}
\put(1416.0,724.0){\rule[-0.200pt]{4.818pt}{0.400pt}}
\put(220.0,801.0){\rule[-0.200pt]{4.818pt}{0.400pt}}
\put(198,801){\makebox(0,0)[r]{0.03878}}
\put(1416.0,801.0){\rule[-0.200pt]{4.818pt}{0.400pt}}
\put(220.0,877.0){\rule[-0.200pt]{4.818pt}{0.400pt}}
\put(198,877){\makebox(0,0)[r]{0.0388}}
\put(1416.0,877.0){\rule[-0.200pt]{4.818pt}{0.400pt}}
\put(220.0,113.0){\rule[-0.200pt]{0.400pt}{4.818pt}}
\put(220,68){\makebox(0,0){0}}
\put(220.0,857.0){\rule[-0.200pt]{0.400pt}{4.818pt}}
\put(372.0,113.0){\rule[-0.200pt]{0.400pt}{4.818pt}}
\put(372,68){\makebox(0,0){2}}
\put(372.0,857.0){\rule[-0.200pt]{0.400pt}{4.818pt}}
\put(524.0,113.0){\rule[-0.200pt]{0.400pt}{4.818pt}}
\put(524,68){\makebox(0,0){4}}
\put(524.0,857.0){\rule[-0.200pt]{0.400pt}{4.818pt}}
\put(676.0,113.0){\rule[-0.200pt]{0.400pt}{4.818pt}}
\put(676,68){\makebox(0,0){6}}
\put(676.0,857.0){\rule[-0.200pt]{0.400pt}{4.818pt}}
\put(828.0,113.0){\rule[-0.200pt]{0.400pt}{4.818pt}}
\put(828,68){\makebox(0,0){8}}
\put(828.0,857.0){\rule[-0.200pt]{0.400pt}{4.818pt}}
\put(980.0,113.0){\rule[-0.200pt]{0.400pt}{4.818pt}}
\put(980,68){\makebox(0,0){10}}
\put(980.0,857.0){\rule[-0.200pt]{0.400pt}{4.818pt}}
\put(1132.0,113.0){\rule[-0.200pt]{0.400pt}{4.818pt}}
\put(1132,68){\makebox(0,0){12}}
\put(1132.0,857.0){\rule[-0.200pt]{0.400pt}{4.818pt}}
\put(1284.0,113.0){\rule[-0.200pt]{0.400pt}{4.818pt}}
\put(1284,68){\makebox(0,0){14}}
\put(1284.0,857.0){\rule[-0.200pt]{0.400pt}{4.818pt}}
\put(1436.0,113.0){\rule[-0.200pt]{0.400pt}{4.818pt}}
\put(1436,68){\makebox(0,0){16}}
\put(1436.0,857.0){\rule[-0.200pt]{0.400pt}{4.818pt}}
\put(220.0,113.0){\rule[-0.200pt]{292.934pt}{0.400pt}}
\put(1436.0,113.0){\rule[-0.200pt]{0.400pt}{184.048pt}}
\put(220.0,877.0){\rule[-0.200pt]{292.934pt}{0.400pt}}
\put(45,495){\makebox(0,0){${\tilde{R}}$}}
\put(828,23){\makebox(0,0){Order of pert.theory}}
\put(220.0,113.0){\rule[-0.200pt]{0.400pt}{184.048pt}}
\put(372,438){\raisebox{-.8pt}{\makebox(0,0){$\star$}}}
\put(448,243){\raisebox{-.8pt}{\makebox(0,0){$\star$}}}
\put(524,365){\raisebox{-.8pt}{\makebox(0,0){$\star$}}}
\put(600,330){\raisebox{-.8pt}{\makebox(0,0){$\star$}}}
\put(676,348){\raisebox{-.8pt}{\makebox(0,0){$\star$}}}
\put(752,340){\raisebox{-.8pt}{\makebox(0,0){$\star$}}}
\put(828,345){\raisebox{-.8pt}{\makebox(0,0){$\star$}}}
\put(904,342){\raisebox{-.8pt}{\makebox(0,0){$\star$}}}
\put(980,343){\raisebox{-.8pt}{\makebox(0,0){$\star$}}}
\put(1056,343){\raisebox{-.8pt}{\makebox(0,0){$\star$}}}
\put(1132,343){\raisebox{-.8pt}{\makebox(0,0){$\star$}}}
\put(1208,343){\raisebox{-.8pt}{\makebox(0,0){$\star$}}}
\put(1284,343){\raisebox{-.8pt}{\makebox(0,0){$\star$}}}
\put(1360,343){\raisebox{-.8pt}{\makebox(0,0){$\star$}}}
\put(372,343){\usebox{\plotpoint}}
\put(372.0,343.0){\rule[-0.200pt]{238.009pt}{0.400pt}}
\end{picture}
\caption{As Fig 5, but at ${\sqrt{s}}={M_Z}$
}
\label{fig:Figure6}
\end{center}
\end{figure}
{\hspace{-.7cm}Clearly}
at the higher energy the agreement is much improved. With the fixed-order results 
exactly tracking the all-orders result for $n>4$. Wild oscillations only set in for
$n>30$ at this higher energy. \\
Finally, in Figure 7 we show a plot of ${\tilde{R}}(s)$  versus ${\ln}({\sqrt{s}}/{\rm{GeV}})$,
in the range $1<{\sqrt{s}}<91$ GeV. The solid line corresponds to the all-orders resummed
result and the dashed line to the NNLO fixed-order CORGI result. We assume ${\alpha}_{s}({M}_{Z}^{2})=0.119$,
and evolve through flavour thresholds using the three-loop matching condition \cite{r32,r33}.\\
The QED fine structure constant is extremely well-measured with
\be
{\alpha}^{-1}{\equiv}{{\alpha}(0)}^{-1}=137.03599976(50)\;.
\ee
If one wishes to evolve ${\alpha}$ away from $s=0$ to obtain ${\alpha}(s)$, however, then
one needs to know leptonic and hadronic corrections,
\be
{{\alpha}(s)}^{-1}=(1-{\Delta}{\alpha}_{\rm{lep}}(s)-{\Delta}{\alpha}_{\rm{had}}(s)-
{\Delta}{\alpha}_{\rm{top}}(s)){\alpha}^{-1}\;.
\ee
Whilst the leptonic corrections are known at three loops  and are well-determ-
\newline
ined \cite{r35},
the hadronic corrections for the contribution of {\hspace{.2cm}the} {\hspace{.2cm}five} {\hspace{.2cm}lightest}
\newpage 
\begin{figure}
\begin{center}
\input{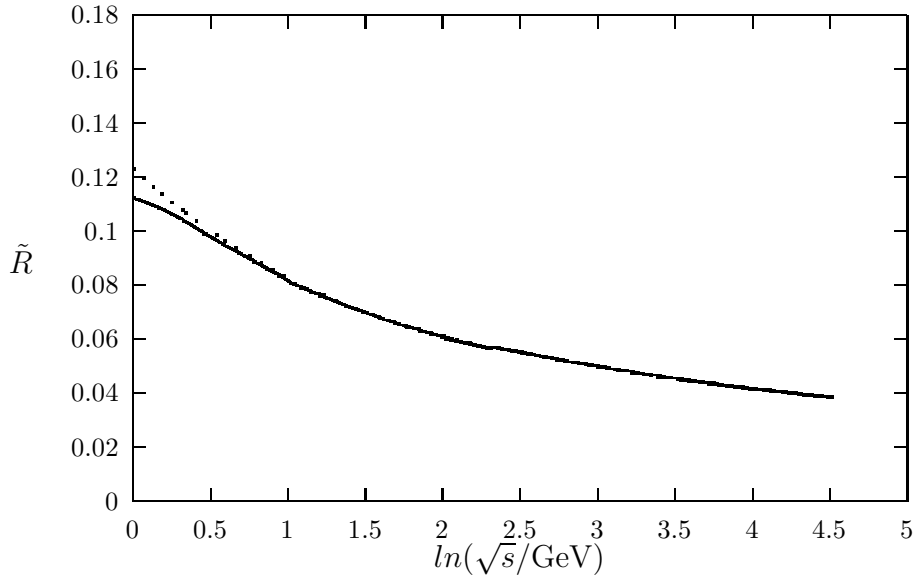}
\caption{
Fixed order (dashed line) and all-orders renormalon resummations (solid line) for ${\tilde{R}}(s)$
versus $ln({\sqrt{s}}/{\rm{GeV}})$, over the range $1<{\sqrt{s}}<91$ GeV.
}
\label{fig:Figure7}
\end{center}
\end{figure}
{\hspace{-.7cm}flavours},
which we
have denoted ${\Delta}{\alpha}_{\rm{had}}(s)$, is rather poorly determined and has to
be reconstructed from the $s$-dependence of ${\tilde{R}}(s)$ using a dispersion relation.
The contribution of the heaviest flavour ${\Delta}{\alpha}_{\rm{top}}(s)$ is rather well-determined
and can be included separately. The value of ${\alpha}({M^2_Z})$ is of particular
relevance since it limits the precision with which the unknown Higgs mass $M_H$ of the
Standard Model can be predicted from precision electroweak corrections \cite{r34}. Taking
${s}={M}_{Z}^{2}$ we have \cite{r35} ${\Delta}{\alpha}_{\rm{lep}}({M}_{Z}^{2})=314.98{\times}{10}^{-4}$
and ${\Delta}{\alpha}_{\rm{top}}({M}_{Z}^{2})=-0.76{\times}{10}^{-4}$.  For 
{\hspace{.0006cm}the} {\hspace{.01cm}hadronic
}
{\hspace{.10cm}contribution} {\hspace{.10cm}we} {\hspace{.10cm}can} {\hspace{.10cm}use} 
{\hspace{.10cm}the
}
{\hspace{-.75cm}dispersion} relation,
\be
{\Delta}{\alpha}_{\rm{had}}({M}_{Z}^{2})=-\frac{{\alpha}{M}_{Z}^{2}}{3{\pi}}{PV}{\int_{4{m}_{\pi}^{2}}^
{\infty}}{ds}\frac{R(s)}{s(s-{M}_{Z}^{2})}\;.
\ee
In Ref.\cite{r24} new exclusive data from BES-II \cite{r36} and Novosibirsk \cite{r37} have been used to extract
${R}(s)$ in the low energy region, with NNLO fixed order perturbative QCD used to
evaluate it in the ranges $2.8<{\sqrt{s}}<3.74$ and $5<{\sqrt{s}}<{\infty}$. We plan to
approximate $R(s)$ in these latter ranges by the all-orders and  NNLO fixed-order results
for $R(s)$, as plotted in Figure 7. We shall use the exclusive data results as in Ref.\cite{r24}, in
the remaining energy ranges. Taking ${\alpha}_{s}(M^2_z)=0.119$ we shall then determine
${\alpha}({M}_{Z}^{2})$ from the fixed-order CORGI results, and the all-orders
leading-$b$ resummed results. Since these results are contour-improved they include
a resummation of analytical continuation terms not included in the fixed-order
perturbative results used in \cite{r24}. We are interested in establishing if these
terms and the uncalculated higher-order corrections, as estimated by the leading-$b$
approximation, cause a significant shift in ${\alpha}({M^2_{Z}})$, and whether
this has any ramifications for the constraints on ${M}_{H}$. In the region
$2.8<{\sqrt{s}}<3.74$ we obtain
 ${\Delta}{{\alpha}_{\rm{had}}}({M^2_{Z}})=(9.5424{\times}{10}^{-4}$,
9.7075${\times}{10}^{-4})$ for the (fixed-order, all-orders) CORGI results, and in the
region $5<{\sqrt{s}}<{\infty}$ we find ${\Delta}{\alpha}_{\rm{had}}({M^2_{Z}})=(170.788{\times}{10}^{-4}$,
170.635${\times}{10}^{-4})$. We find correspondingly using Eq.(41), ${{\alpha}({M^2_{Z}})}^{-1}=(128.967,128.971)$, 
to be compared with ${{\alpha}({M^2_{Z}})}^{-1}=128.978{\pm}0.027$ quoted in Ref \cite{r24}. We conclude
that the analytical continuation terms and uncalculated higher order perturbative corrections
do not cause a significant change in ${\alpha}({M^2_{Z}})$, and their inclusion does
nothing modify the conclusions of Ref \cite{r24}.
\section*{6 All-orders CORGI resummations for the scalar correlator}
The Higgs decay width to a quark anti-quark pair will be of fundamental phenomenological
importance. In practice the decay to a $b{\bar{b}}$ will be the dominant contribution
\be
{\Gamma}(H{\rightarrow}{b}{\overline{b}})=\frac{3{G_F}}{4\sqrt{2}{\pi}}
{M_H}{m}_{b}^{2}({M}_{H}^{2}){\cal{R}}({M}_{H}^{2})\;. 
\ee
Here $M_H$ is the Higgs mass and ${m_b}({M}_{H}^{2})$ is the running $b$-quark mass.
${\cal{R}}({M}_{H}^{2})$ is a coefficient function 
with a perturbative expansion
\be
{\cal{R}}({M}_{H}^{2})= 1+{\sum_{n>0}}{r}_{n}{a}^{n}\;,
\ee
where the coefficients ${r_1},{r_2},{r_3}$ have been exactly computed \cite{r38}. 
${\cal{R}}$ can be straightforwardly related to the scalar correlator ${\Pi}_{s}(s)$.
 One can define an analogue of the vector Adler $D$-function so that (cf. Eq.(8))
\be
D(s)=s\frac{d}{ds}\left[\frac{{\Pi}_{s}(s)}{s}\right]\;,
\ee
This may be written in terms of the coefficient function ${\cal{D}}(s)$ where
\be
D(s)=\frac{3}{8{\pi}^{2}}{({m}_{b}(s))}^{2}{\cal{D}}(s)\;,
\ee
and ${\cal{D}}$ has the perturbative expansion, 
\be
{\cal{D}}(s)=1+{\sum_{n>0}}{d_n}{a}^{n}.
\ee
${m}_{b}^{2}(s){\cal{R}}(s)$ can be related to ${m}_{b}^{2}(-s){\cal{D}}(-s)$ by
analytical continuation from Euclidean to Minkowski, and one can write a
representation of the same form as Eq.(10) with
\be
{m}_{b}^{2}({M}_{H}^{2})\;{\cal{R}}({M}_{H}^{2})=\frac{1}{2{\pi}}
{\int_{-{\pi}}^{\pi}}{d{\theta}}\; {m}_{b}^{2}({e}^{i{\theta}}{M}_{H}^{2})
{\cal{D}}({e}^{i{\theta}}{M}_{H}^{2})\;.
\ee
To proceed further we can express the running mass in terms of an RG-invariant
mass ${\hat{m}}_{b}$ and the mass anomalous dimension ${\gamma}_{m}(a)$, defined by
\be
-\frac{d{{\ln}{\left(m(s)\right)}}}{d{\ln}({s})}={\gamma}_{m}(a)={\sum_{i{\geq}0}}{\gamma}_{i}{a}^{i+1}\;.
\ee
We can then write \cite{r21}
\be
{{m}_{b}}^{2}(s)={\hat{m}_{b}}^{2}{b}^{\frac{4{{\gamma}_{0}}}{b}}
{\left(\frac{a}{1+ca}\right)}^{\frac{4{{\gamma}_{0}}}{b}}
{\exp}\left[4{{\int}_{0}^{a}}{dx}
\frac{{\gamma}_{1}+({\gamma}_{1}c+{\gamma}_{2}-{\gamma}_{0}{c_2})x+{\ldots}}
{b(1+cx)(1+cx+{c_2}{x}^{2}+{\ldots})}\right]\;.
\ee
The ${b}^{\frac{4{\gamma}_{0}}{{b}}}$ is the standard normalization of the definition of the RG-invariant
mass ${\hat{m}}_{b}$.
One can then write a CORGI series for ${m}_{b}^{2}(s)$ ${\cal{D}}(s)$ exactly equivalent to
that for the moments of structure functions in Ref.\cite{r21},
\be
{{m}_{b}^{2}}(s){\cal{D}}(s)=
{\hat{m}_{b}}^{2}{b}^{\frac{4{{\gamma}_{0}}}{b}}
{\left(\frac{{{a}_{0}}(s)}{1+c{{a}_{0}}(s)}
\right)}^{\frac{4{{\gamma}_{0}}}{b}}(1+{X}_{2}{a}_{0}^{2}(s)+{X}_{3}{a}_{0}^{3}(s)+{\ldots}+{X}_{n}{a}_{0}^{n}(s)
+{\ldots})\;
\ee
where ${a}_{0}(s)$ denotes the CORGI coupling which is again defined in terms of the Lambert $W$-function
as in Eq.(17), and
with the anomalous dimension present one now has
 ${\Lambda}_{D}={\exp}[(\frac{{\gamma}_{1}}{{\gamma}_{0}{b}})
+(\frac{d}{4{\gamma}_{0}})]
{(\frac{2c}{b})}^{-\frac{c}{b}}{\Lambda}_{\overline{MS}}$ , with $d$ the coefficient $d_1$ in the
${\overline{MS}}$ factorization
and renormalization scheme with ${M}^{2}={\mu}^{2}=s$ ($M$ denoting
the factorization scale) \cite{r21}. The exactly known CORGI invariants $X_2$ and $X_3$ follow from
Eqs.(18) of Ref.\cite{r21} , and allowing for the different definition of the anomalous dimension one needs
to replace $d_i$ in Ref.\cite{r21} by $4{\gamma}_i$ . Lumping various inessential prefactors together we
can define
\be
{\Gamma}(H{\rightarrow}{b}{\bar{b}})=\frac{3{G_{F}}}{4{\sqrt{2}}{\pi}}{M_H}{\hat{m}}_{b}^{2}{b}^
{4{\gamma}_{0}/{b}}{\Gamma}\;,
\ee
where ${\Gamma}$ has the contour-improved CORGI representation,
\be
{\Gamma}= \frac{1}{2{\pi}}{\int_{-{\pi}}^{\pi}}{d{\theta}}{\left(\frac{{\bar{a}}_{0}}{1+c{\bar{a}}_{0}}
\right)}^{4{\gamma}_{0}/{b}}(1+{X}_{2}{\bar{a}}_{0}^{2}+{X}_{3}{\bar{a}}_{0}^{3}+
{\sum_{n>3}}{X}_{n}^{(L)}{\bar{a}}_{0}^{n})\;,
\ee
with ${\bar{a}}_{0}={{a}_{0}}({e^{i\theta}}{M^2_H})$.\\
In the scalar case one will have coefficients with the structure
\be
{d_n}={d}_{n}^{[n-1]}{N}_{f}^{n-1}+{d}_{n-1}^{[n-1]}{N}_{f}^{n-2}+{\ldots}+{d}_{n}^{[0]}\;,
\ee
and after replacing ${N_f}=({\frac{33}{2}}-3b)$, as before, one arrives at a leading-$b$ term with
the structure ${d^L_{n}}$=${(-3)}^{n-1}{d^{[n-1]}_{n}}{b}^{n-1}$, with one less power
of $b$. The anomalous dimension coefficients will have the structure ${\gamma}_{n}^{(L)}=
{\gamma}_{n}^{(n)}{b}^{n}$, but since the anomalous dimension ${\gamma}_{m}(a)$ does not
contain renormalons there is no motivation for making this approximation, and it is
poor in practice, as noted in Ref.\cite{r24a}. Whilst an all-orders result for ${\gamma}_{n}^{(L)}$
does exist \cite{r24a}, we shall follow Ref.\cite{r24a} and set ${\gamma}_{n}^{(L)}=0$ for $n>0$, retaining
only ${\gamma}_{0}$. The all-orders result for ${X}_{n}^{(L)}$ follow straightforwardly
from ${d}_{n}^{(L)}$. From the large-$N_f$ results of Ref.\cite{r24a} for the scalar correlator
one can obtain an explicit all-orders expression for ${d}_{n}^{(L)}(V)$ (in the V-scheme)
analogous to Eq.(21) in the vector case. For $n$ even one has,
\ba
{d}_{n}^{(L)}(V)&=&-\frac{32}{3}\frac{1}{{2}^{n+1}}\left(1-\frac{1}{{2}^{n}}\right){\zeta}(n+1)
{n!}{b}^{n-1}+\left(\frac{4}{n}+\frac{4}{3}\right){\left(\frac{1}{2}\right)}^{n-1}{n!}{b}^{n-1}
\nonumber\\
&&-\left({\frac{1}{n}}
+{\frac{1}{3}}\right)
{\left(\frac{1}{4}\right)}^{n-1}{n!}{b}^{n-1}\;,
\ea
whilst for odd $n$ one has,
\be
{d}_{n}^{(L)}(V)=\left(\frac{4}{n}+\frac{4}{3}\right){\left(\frac{1}{2}\right)}^{n-1}
{n!}{b}^{n-1}-\left(\frac{1}{n}+\frac{1}{3}\right){\left(\frac{1}{4}\right)}^{n-1}{n!}{b}^{n-1}\;.
\ee
As in the vector case one can define a leading-$b$ resummation
\be
{\cal{D}}^{(L)}=1+{\sum_{k=1}^{\infty}}{d}_{k}^{(L)}(V){a}^{k}\;,
\ee
analogous to Eq.(22), which may be defined as a regulated Borel sum
\be
{\cal{D}}^{(L)}(1/a)= 1 + PV{{\int_{0}^{\infty}}\;{dz}\;{e}^{-z/a}[{G}_{-}(z)+{G}_{+}
(z)]}\;.
\ee
Here ${G}_{-}(z)$ and ${G}_{+}(z)$ are the contributions to the Borel transform from
UV and IR renormalons, respectively. One has (in the V-scheme) \cite{r24a}
\ba
{G}_{-}(z)&=&\;-\frac{4}{3}{\sum_{k=1}^{\infty}}\frac{{(-1)}^k}{{k^2}{(1+bz/2k)}
^{2}}
\nonumber \\
{G}_{+}(z)&=& \; \frac{4}{(1-bz/2)}\;-\;\frac{1}{(1-bz/4)}\;+\frac{4}{3}
{\sum_{k=3}^{\infty}}
\frac{{(-1)}^{k}}{{k}^{2}{(1-bz/2k)}^{2}}\;
\ea
>From these expressions one can read off the residues $A_0$, ${B}_{0}$ (cf. Eq.(24)),
and one can then calculate ${\cal{D}}^{(L)}(F){{\mid}_{UV}}$ and ${\cal{D}}^{(L)}(F){{\mid}_{IR}}$
using Eqs.(27),(28). Finally
\be
{\cal{D}}^{(L)}(F)={\cal{D}}^{(L)}(F){{\mid}_{UV}}+{\cal{D}}^{(L)}(F){{\mid}_{IR}}\;,
\ee
with $F{\equiv}{1/{a_{V}}}$.
To perform the leading-$b$ CORGI resummation in Eq.(53) we simply
need to relate ${a}_{V}^{(L)}$ and ${a}_{0}^{(L)}$ as we did in Section 3. In the presence
of the anomalous dimension the RS-invariant combination ${\rho}_{0}$ in Eq.(32) is
replaced by the factorization scheme and RS (FRS) invariant combination $X_1$ introduced
in Ref.\cite{r21},
\be
{X}_{1}=4{\gamma}_{0}{\ln}\left(\frac{M}{\tilde{\Lambda}}\right)-\frac{4{\gamma}_{1}}{b}-{d}_{1}(M)\;,
\ee
where $M$ is the factorization scale. Recalling that we have decided to set ${\gamma}_{i}^{(L)}=0$ for
$i>0$ in our leading-$b$ resummations , we can use Eq.(61) to relate ${a}^{(L)}_{0}$ and
${a}_{V}^{(L)}$,
\be
\frac{1}{{a}_{V}^{(L)}}=\frac{1}{{a}_{0}^{(L)}}+\frac{b}{4{\gamma}_{0}}{d}_{1}^{(L)}
(V)\;.
\ee
\newpage
\begin{figure}
\begin{center}
\setlength{\unitlength}{0.240900pt}
\ifx\plotpoint\undefined\newsavebox{\plotpoint}\fi
\sbox{\plotpoint}{\rule[-0.200pt]{0.400pt}{0.400pt}}%
\begin{picture}(1500,900)(0,0)
\font\gnuplot=cmr10 at 10pt
\gnuplot
\sbox{\plotpoint}{\rule[-0.200pt]{0.400pt}{0.400pt}}%
\put(220.0,164.0){\rule[-0.200pt]{4.818pt}{0.400pt}}
\put(198,164){\makebox(0,0)[r]{0.0416}}
\put(1416.0,164.0){\rule[-0.200pt]{4.818pt}{0.400pt}}
\put(220.0,266.0){\rule[-0.200pt]{4.818pt}{0.400pt}}
\put(198,266){\makebox(0,0)[r]{0.0418}}
\put(1416.0,266.0){\rule[-0.200pt]{4.818pt}{0.400pt}}
\put(220.0,368.0){\rule[-0.200pt]{4.818pt}{0.400pt}}
\put(198,368){\makebox(0,0)[r]{0.042}}
\put(1416.0,368.0){\rule[-0.200pt]{4.818pt}{0.400pt}}
\put(220.0,470.0){\rule[-0.200pt]{4.818pt}{0.400pt}}
\put(198,470){\makebox(0,0)[r]{0.0422}}
\put(1416.0,470.0){\rule[-0.200pt]{4.818pt}{0.400pt}}
\put(220.0,571.0){\rule[-0.200pt]{4.818pt}{0.400pt}}
\put(198,571){\makebox(0,0)[r]{0.0424}}
\put(1416.0,571.0){\rule[-0.200pt]{4.818pt}{0.400pt}}
\put(220.0,673.0){\rule[-0.200pt]{4.818pt}{0.400pt}}
\put(198,673){\makebox(0,0)[r]{0.0426}}
\put(1416.0,673.0){\rule[-0.200pt]{4.818pt}{0.400pt}}
\put(220.0,775.0){\rule[-0.200pt]{4.818pt}{0.400pt}}
\put(198,775){\makebox(0,0)[r]{0.0428}}
\put(1416.0,775.0){\rule[-0.200pt]{4.818pt}{0.400pt}}
\put(220.0,877.0){\rule[-0.200pt]{4.818pt}{0.400pt}}
\put(198,877){\makebox(0,0)[r]{0.043}}
\put(1416.0,877.0){\rule[-0.200pt]{4.818pt}{0.400pt}}
\put(220.0,113.0){\rule[-0.200pt]{0.400pt}{4.818pt}}
\put(220,68){\makebox(0,0){2}}
\put(220.0,857.0){\rule[-0.200pt]{0.400pt}{4.818pt}}
\put(372.0,113.0){\rule[-0.200pt]{0.400pt}{4.818pt}}
\put(372,68){\makebox(0,0){4}}
\put(372.0,857.0){\rule[-0.200pt]{0.400pt}{4.818pt}}
\put(524.0,113.0){\rule[-0.200pt]{0.400pt}{4.818pt}}
\put(524,68){\makebox(0,0){6}}
\put(524.0,857.0){\rule[-0.200pt]{0.400pt}{4.818pt}}
\put(676.0,113.0){\rule[-0.200pt]{0.400pt}{4.818pt}}
\put(676,68){\makebox(0,0){8}}
\put(676.0,857.0){\rule[-0.200pt]{0.400pt}{4.818pt}}
\put(828.0,113.0){\rule[-0.200pt]{0.400pt}{4.818pt}}
\put(828,68){\makebox(0,0){10}}
\put(828.0,857.0){\rule[-0.200pt]{0.400pt}{4.818pt}}
\put(980.0,113.0){\rule[-0.200pt]{0.400pt}{4.818pt}}
\put(980,68){\makebox(0,0){12}}
\put(980.0,857.0){\rule[-0.200pt]{0.400pt}{4.818pt}}
\put(1132.0,113.0){\rule[-0.200pt]{0.400pt}{4.818pt}}
\put(1132,68){\makebox(0,0){14}}
\put(1132.0,857.0){\rule[-0.200pt]{0.400pt}{4.818pt}}
\put(1284.0,113.0){\rule[-0.200pt]{0.400pt}{4.818pt}}
\put(1284,68){\makebox(0,0){16}}
\put(1284.0,857.0){\rule[-0.200pt]{0.400pt}{4.818pt}}
\put(1436.0,113.0){\rule[-0.200pt]{0.400pt}{4.818pt}}
\put(1436,68){\makebox(0,0){18}}
\put(1436.0,857.0){\rule[-0.200pt]{0.400pt}{4.818pt}}
\put(220.0,113.0){\rule[-0.200pt]{292.934pt}{0.400pt}}
\put(1436.0,113.0){\rule[-0.200pt]{0.400pt}{184.048pt}}
\put(220.0,877.0){\rule[-0.200pt]{292.934pt}{0.400pt}}
\put(45,495){\makebox(0,0){${\Gamma}$}}
\put(828,23){\makebox(0,0){Order of pert.theory}}
\put(220.0,113.0){\rule[-0.200pt]{0.400pt}{184.048pt}}
\put(296,559){\raisebox{-.8pt}{\makebox(0,0){$\star$}}}
\put(372,469){\raisebox{-.8pt}{\makebox(0,0){$\star$}}}
\put(448,515){\raisebox{-.8pt}{\makebox(0,0){$\star$}}}
\put(524,493){\raisebox{-.8pt}{\makebox(0,0){$\star$}}}
\put(600,504){\raisebox{-.8pt}{\makebox(0,0){$\star$}}}
\put(676,499){\raisebox{-.8pt}{\makebox(0,0){$\star$}}}
\put(752,500){\raisebox{-.8pt}{\makebox(0,0){$\star$}}}
\put(828,502){\raisebox{-.8pt}{\makebox(0,0){$\star$}}}
\put(904,497){\raisebox{-.8pt}{\makebox(0,0){$\star$}}}
\put(980,505){\raisebox{-.8pt}{\makebox(0,0){$\star$}}}
\put(1056,494){\raisebox{-.8pt}{\makebox(0,0){$\star$}}}
\put(1132,509){\raisebox{-.8pt}{\makebox(0,0){$\star$}}}
\put(1208,488){\raisebox{-.8pt}{\makebox(0,0){$\star$}}}
\put(1284,514){\raisebox{-.8pt}{\makebox(0,0){$\star$}}}
\put(1360,486){\raisebox{-.8pt}{\makebox(0,0){$\star$}}}
\sbox{\plotpoint}{\rule[-0.400pt]{0.800pt}{0.800pt}}%
\put(296,501){\usebox{\plotpoint}}
\put(296.0,501.0){\rule[-0.400pt]{256.318pt}{0.800pt}}
\end{picture}
\caption{Fixed-order CORGI results for ${\Gamma}$ (${M_H}=115$ GeV) in ${\rm{N}}^{n}$LO perturbation
theory (starred points), compared to the all-orders resummation (solid line)
}
\label{fig:Figure8}
\end{center}
\end{figure}
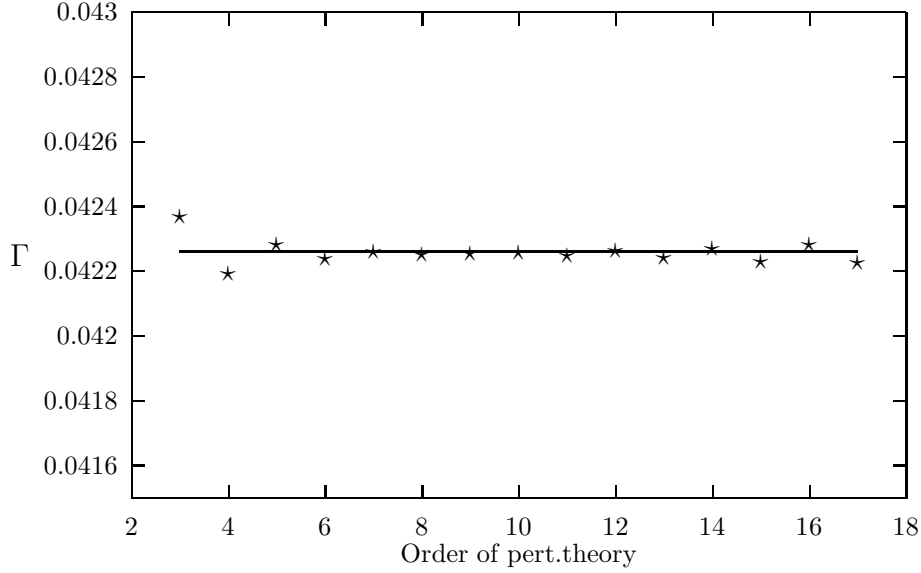
It then follows that the all-orders formal resummation in Eq.(53) is given by
\ba
&&{\Gamma}=\frac{1}{2{\pi}}{\int_{-{\pi}}^{\pi}}{d{\theta}}{\left(\frac{{\bar{a}}_{0}}{1+c{\bar{a}}_{0}}
\right)}^{4{\gamma}_{0}/b}
[1-\frac{4{\gamma}_{0}}{b}{\ln}\left(1+{\frac{4{\gamma}_{0}}{b}}{d}_{1}^{(L)}(V)
{\bar{a}}_{0}\right)
\nonumber\\
&&+{\cal{D}}^{(L)}\left(\frac{1}{{\bar{a}}_{0}}+\frac{b}{4{\gamma}_{0}}{d}_{1}^{(L)}(V)
\right)
+({X}_{2}-{X}_{2}^{(L)}){\bar{a}}_{0}^{2}+({X}_{3}-{X}_{3}^{(L)}){\bar{a}}_{0}^{3}]\;.
\ea
The logarithm term arises because of the fractional power ${a}^{4{\gamma}_{0}/b}$. Relating
the $V$-scheme and CORGI couplings at the leading-$b$ level one has,
\be
{{a}_{V}^{(L)}}^{4{\gamma}_{0}/b}={{a}_{0}^{(L)}}^{4{\gamma}_{0}/b}
{\left(1+{\frac{4\gamma_{0}}{b}}{d_{1}}^{(L)}(V){a_{0}}^{(L)}\right)}
^{{-4\gamma_{0}}/b}\;.
\ee
On expanding using the binomial theorem only the terms {\it linear} in ${\gamma}_{0}$ are
leading in $b$, the remainder should be discarded. Writing the binomial expansion
as ${\exp}[(-4{\gamma}_{0}/b){\ln}S]=1-(4{\gamma}_{0}/b){\ln}S+O({{\gamma}_{0}}^{2})$, the
result follows. The same subtlety enters in deriving an analogue of Eq.(36) to
generate the ${X}_{n}^{(L)}$ in terms of ${d}_{n}^{(L)}(V)$ explicitly given by Eqs.(55),(56).
One finds
\be
{X}_{n}^{(L)}={\cal{C}}_{n}\left[{\sum_{k=1}^{\infty}}{d}_{n}^{(L)}(V)
{\left(\frac{a}{{1+(4{\gamma}_{0}/b)}{d}_{1}^{(L)}(V)a}\right)}^{k}
-{\frac{4{\gamma}_{0}}{b}}
ln\left(1+{\frac{4\gamma_0}{b}}{d_1}^{(L)}(V)a\right)\right].\
\ee
These results can then be used to calculate all-orders and fixed-order CORGI predictions. We
give in Figure 8 the analogue of Figs.1,5,6 , for the Higgs decay width (with pre-factor
set to unity) ${\Gamma}$, we set ${M_H}=115$ GeV and ${\alpha}_{s}({M}_{Z}^{2})=0.119$.
The starred points show the fixed-order CORGI results, and the solid line the all-orders
resummation. As before the agreement of the highest exactly calculated $n=3$ fixed-order
result with the all-orders result is good. The fixed-order results track the resummed
result up to $n=12$, beyond which an oscillatory trend is noticeable. From Eq.(59) one
can see that ${UV}_{1}$ and ${IR}_{1}$ renormalon singularities are present, and
so the leading asymptotics are not dominated by ${UV}_{1}$ as in the vector case. The
process of analytical continuation , however, serves to remove ${IR}_{1}$ \cite{r7,r8} and so the
leading asymptotics of ${\Gamma}$ is expected to be dominated by the leading ${UV}_{1}$
renormalon, with resulting alternating-sign factorial behaviour. As discussed in Ref.\cite{r24a}
the presence of the leading ${IR}_{1}$ renormalon in ${\cal{D}}$ suggests that the
obvious generalization of the Adler function in Eq.(45) may not be optimal, and an 
alternative is suggested. For our purposes here we are simply using $D$ as a tool
to compute the physical quantity ${\Gamma}$, and so this is not a problem.
\section*{7 Analytic expressions for the CORGI contour improvement}
In this section we wish to point out that we can obtain explicit
analytic expressions for the CORGI fixed-order contour improved results
for ${\tilde{R}}(s)$ in terms of the Lambert $W$-function, eliminating the
need for numerical Simpson's rule evaluation. From Eq.(17) we see that
we can write the CORGI coupling ${\bar{a}}_{0}={a}_{0}({e}^{i{\theta}}s)$ in terms of
the Lambert $W$-function as,
\be
{\bar{a}}_{0}=\frac{-1}{c[1+W(A{e}^{iK{\theta}})]}\;,
\ee
where 
\be
A(s)=\frac{-1}{e}{\left(\frac{{\sqrt{s}}}{{\Lambda}_{D}}\right)}^{-b/c}\;\;,K=\frac{-b}{2c}\;.
\ee
Thus after the contour integration the $X_{n-1}$ coefficient multiplies
\ba
{A}_{n}(s)&{\equiv}&\frac{1}{2{\pi}}{\int_{-{\pi}}^{\pi}}{d{\theta}}{{\bar{a}}_{0}}^{n}
=\frac{1}{2{\pi}}{\int_{-\pi}^{0}}d{\theta}
\frac{(-1)^n}{c^n}{[1+{W}_{1}(A(s)
{{e}^{iK{\theta}})]}^{n}}
\nonumber \\
&&+\frac{1}{2{\pi}}
{\int_{0}^{\pi}}{d{\theta}}\frac{{(-1)}^{n}}{{c}^{n}}{[1+{W}_{-1}(A(s)
{{e}^{iK{\theta}})]}^{n}}\;,
\ea
where the appropriate branches of the $W$-function are to be used in the two regions of
integration. By making the change of variable $w=W(A(s){e}^{iK{\theta}})$ we can then obtain
the above integrals in the form,
\be
\frac{(-1)^{n}}{2iK{c}^{n}{\pi}}{\int}\frac{dw}{{w(1+w)}^{n-1}}\;.
\ee
The $w$-integral is elementary, and including the limits of integration, and noting that
${W}_{1}(A(s){e}^{-iK{\pi}})={[{W}_{-1}(A(s){e}^{iK{\pi}})]}^{*}$, we obtain the explicit result,
\be
{A}_{n}(s)= \frac{{(-1)}^{n}}{{c}^{n}K{\pi}}
Im\left[{\ln}\left(\frac{{W}_{-1}(A(s){e}^{iK{\pi}})}{1+{W}_{-1}
(A(s){e}^{ik{\pi}}}\right)
+{\sum_{k=1}^{n-2}}
\frac{1}{k{(1+{W}_{-1}{A(s){e}^{iK{\pi}})}}^{k}}\right]\;,
\ee
for $n>2$. For n=1 we have ${A}_{1}(s)=(-1/({\pi}Kc))
Im[{\ln}{W}_{-1}(A(s){e}^{iK{\pi}})]$.
 We finally obtain the CORGI contour-improved fixed-order results
in the form,
\be
{\tilde{R}}(s)={A}_{1}(s)+{\sum_{k=2}^{\infty}}
{X_k}{A}_{k+1}(s)\;.
\ee\\

In the scalar correlator case analytic results can also be obtained. The
$X_n$ coefficient in the CORGI series for ${\Gamma}$ in Eq.(53) will multiply
\ba
{A}_{n}({M}_{H}^{2})&{\equiv}&\frac{1}{2{\pi}}{\int_{-\pi}^{\pi}}{d{\theta}}
\left(\frac{{\bar{a}}_{0}}{1+c{\bar{a}}_{0}}\right)^{4{\gamma}_{0}/b}{\bar{a}}_{0}^{n}
\nonumber \\
&&=\frac{1}{2{\pi}}{\int_{-\pi}^{0}}{d{\theta}}
\frac{(-1)^{n}}{{[-{W}_{1}(A{e}^{iK{\theta}})]}^
{4{\gamma_{0}}/b}}\frac{c^{-n-{4\gamma_0}/b}}{{[1+{W}_{1}(A{e}^{iK{\theta}})]}^{n}}
\nonumber\\
&&+\frac{1}{2{\pi}}{\int_{0}^{\pi}}{d{\theta}}
\frac{(-1)^{n}}{{[-{W}_{-1}(A{e}^{iK{\theta}})]}^
{4{\gamma_{0}}/b}}
\frac{c^{-n-{4\gamma_0}/b}}{{[1+{W}_{-1}(A{e}^{iK{\theta}})]}^{n}}\;.
\ea
Here $A=A({M}_{H}^{2})$.
Making the change of variable ${\omega}=-W(A{e}^{iK{\theta}})$ one can then obtain the
above integrals in the form,
\be
\frac{(-1)^{n}}{2{\pi}iK{c}^{n+4{\gamma_0}/b}}
{\int}{d\omega}\frac{{\omega}^{(-4{\gamma_0}/b)-1}}
{{(1-{\omega})}^{n-1}}\;.
\ee
These integrals may be evaluated in terms of the Hypergeometric function $F(a,b;c;z)$ \cite{r40}.
Inserting the limits of integration we obtain the explicit result for $n>0$,
\ba
{A}_{n}({M}_{H}^{2})&=&\frac{-{(-1)}^{n}b}{4{\pi}K{c}^{n+4{\gamma_0}/b}\gamma_0}
Im\left\{{\left[{W_{-1}}(Ae^{iK\pi})\right]}^{-\frac{4\gamma_0}{b}}\right.
\nonumber\\
&&\left.F\left({n-1},-\frac{4{\gamma}_{0}}{b};1-\frac{4{\gamma}_{0}}{b};{W}_{-1}(A{e}^{iK{\pi}})\right)\right\}\;.
\ea
We can finally write the CORGI contour-improved result in the form
\be
{\Gamma}={A}_{0}({M}_{H}^{2})+{\sum_{k=2}^{\infty}}{X}_{k}{A}_{k}({M}_{H}^{2})\;,
\ee
with 
\be
{A}_{0}({M}_{H}^{2})=\frac{1}{{\pi}K{c}^{4{\gamma}_{0}/{b}}}Im\left[
\frac{b}{4{\gamma}_{0}}{({W}_{-1}(A{e}^{iK{\pi}})}^{-4{\gamma}_{0}/b}+\frac{{({W}_{-1}(A{e}^{iK{\pi}})}
^{1-4{\gamma}_{0}/b}}{(1-\frac{4\gamma_0}{b})}\right]\;.
\ee
\section*{8 Conclusions}
In this paper we have focussed on obtaining exact fixed-order, and leading-$b$ estimated
all-orders results for various inclusive  QCD Minkowski observables , related to the
vector correlator. These could be expressed as a contour integral of the suitably weighted Euclidean
Adler $D(-s)$-function in the complex energy squared plane. ${D}(s)$ is truncated at some
fixed-order and the integral performed numerically as described in Section 2. In this way
contour-improved predictions are obtained, in which an infinite subset of known and
potentially large analytical
continuation terms are resummed to all-orders. By employing the CORGI approach, as discussed
in Section 3, we could further resum to all-orders the complete set of ultraviolet logarithms
involving $s$, which build the $s$-dependence of $D(s)$, avoiding any dependence on an
arbitrary renormalization scale ${\mu}$. The remaining approximation is the missing
higher-order CORGI invariants ${X}_{i}\;(i>2)$, which remain unknown since the perturbative
coefficients have only been calculated to NNLO so far. We used the so-called leading-$b$
approximation to estimate these. Using the exact large-$N_f$ all-orders results for $D(s)$ \cite{r7,r8}
we were able to sum the CORGI series to all-orders in terms of a sum of exponential integral
functions, corresponding to the contributions of the ultraviolet and infrared renormalons in
the Borel plane. This was technically far more straightforward then previous analogous
resummations of the Effective Charge beta-function for $D(s)$ \cite{r18,r19}. By comparing the NNLO
fixed-order CORGI results to the all-orders resummations we estimated the uncertainty in
${\alpha}_{s}({M}_{Z}^{2})$, extracted from experimental measurements of ${R}_{\tau}$, to be
${\delta}{\alpha}_{s}({M}_{Z}^{2}){\approx}0.003$. We also showed that using all-orders
and fixed-order contour-improved CORGI results for ${R}(s)$ gave results for the
hadronic corrections to the QED coupling ${\alpha}({M}_{Z}^{2})$ which did not differ
significantly from those obtained using conventional fixed-order perturbation theory in Ref.\cite{r24}.
Finally we showed how recent all-orders large-$N_f$ results for the scalar correlator
could be used to perform analogous resummations for the Higgs decay width to a heavy quark
pair. We finally noted that the CORGI contour-improvement for the $R$-ratio can be written
in analytic form in terms of the Lambert $W$-function, and for the Higgs width in terms of
hypergeometric functions and the Lambert $W$-function, thus avoiding the need to use a 
numerical Simpson's Rule evaluation.\\

There are many possible investigations to be pursued using these methods. In particular
it would be interesting to investigate hadronic corrections to ${\alpha}({M}_{Z}^{2})$ using
the resummed results for $R(s)$ in lower energy ranges where
conventionally inclusive or exclusive data has been used. The scalar correlator results could
also be used to investigate more carefully the uncertainties in estimates of the bottom
quark mass using Sum Rule techniques.
\section*{Acknowledgements}
We would like to thank Andrei Kataev for many entertaining and informative discussions 
about contour-improvement and all-orders resummations. A.M. acknowledges the financial support
of the Iranian government.


\begin{thebibliography}{99}
\bibitem{r1} K. Schilcher and M.D. Tran, Phys. Rev. {\bf D29} (1984) 570.
\bibitem{r2} F. Le Diberder and A. Pich, Phys. Lett. {\bf B289} (1992) 165.
\bibitem{r3} S.G. Gorishny, A.L. Kataev and S.A. Larin, Phys. Lett. {\bf B259} (1991) 144;
L.R. Surguladze and M.A. Samuel, Phys. Rev. Lett. 66 (1991) 560; 66 (1991) 2416 (E).
\bibitem{r4} K.G. Chetyrkin and J.H. Kuhn, Phys. Lett. {\bf B248} (1990) 359; K.G. Chetyrkin
, R. Harlander and J.H. Kuhn, Nucl. Phys. {\bf B586} (2000) 56.
\bibitem{r5} M. Neubert, Nucl. Phys. {\bf B463} (1996) 511.
\bibitem{r6} G. Grunberg hep-ph/9705290.
\bibitem{r7} C.N. Lovett-Turner and C.J. Maxwell, Nucl. Phys. {\bf B432} (1994) 147.
\bibitem{r8} C.N. Lovett-Turner and C.J. Maxwell, Nucl. Phys. {\bf B452} (1995) 188.
\bibitem{r9} D.J. Broadhurst and A.G. Grozin, Phys. Rev. {\bf D52} (1995) 4082.
\bibitem{r10} M. Beneke and V.M. Braun, Phys. Lett, {\bf B348} (1995) 513.
\bibitem{r11} P. Ball, M. Beneke and V.M. Braun, Nucl. Phys. {\bf B452} (1995) 563.
\bibitem{r12} M. Beneke, Nucl. Phys. {\bf B405} (1993) 424.
\bibitem{r13} D.J. Broadhurst, Z. Phys. {\bf C58} (1993) 339.
\bibitem{r14} D.J. Broadhurst and A.L. Kataev, Phys. Lett {\bf B315} (1993) 179.
\bibitem{r15} Stanley J. Brodsky, Einan Gardi, Georges Grunberg and Johann Rathsman,
hep-ph/0002065.
\bibitem{r16} M. Beneke, V.M. Braun and N. Kivel, Phys. Lett. B404 (1997) 315.
\bibitem{r17} C.J. Maxwell, Phys. Lett. {\bf B409} (1997) 382.
\bibitem{r18} C.J. Maxwell and D.G. Tonge, Nucl. Phys. {\bf B481} (1996) 681.
\bibitem{r19} C.J. Maxwell and D.G. Tonge, Nucl. Phys. {\bf B535} (1998) 19.
\bibitem{r20} G. Grunberg, Phys. Lett. {\bf B95} (1980) 70; Phys. Rev. {\bf D29} (1984) 2315.
\bibitem{r21} C.J. Maxwell and A. Mirjalili, Nucl. Phys. {\bf B577} (2000) 209.
\bibitem{r22} ALEPH collaboration, Eur. Phys. J {\bf C4} (1998) 409.
\bibitem{r23} M. Girone and M. Neubert, Phys. Rev. Lett. 76 (1996) 3061.
\bibitem{r24} A.D. Martin, J. Outhwaite and M.G. Ryskin, Phys. Lett. {\bf B492} (2000) 69.
\bibitem{r24a} D.J. Broadhurst, A.L. Kataev and C.J. Maxwell, Nucl. Phys. {\bf B592} (2001) 247.
\bibitem{r25} W.J. Marciano and A. Sirlin, Phys. Rev. Lett. 61 (1988) 1815; 56 (1986) 22.
\bibitem{r26} A. Pich, private communication. This approach is used in the computer program
written by F. Le Diberder, widely applied in experimental analyses.
\bibitem{r27} S.J. Burby and C.J. Maxwell, hep-ph/0011203 (2000).
\bibitem{r28} Einan Gardi, Georges Grunberg and Marek Karliner, JHEP 07 (1998) 007.
\bibitem{r29} G. Grunberg, Phys. Lett. {\bf B325} (1994) 441.
\bibitem{r30} S. Peris and E. de Rafael, Nucl. Phys. {\bf B500} (1997) 325.
\bibitem{r31} P.M. Stevenson, Phys. Rev. {\bf D23} (1981) 2916.
\bibitem{r32} K.G. Chetyrkin, B.A. Kniehl and M. Steinhauser, Phys. Rev. Lett. 79 (1997) 2184.
\bibitem{r33} German Rodrigo, Antonio Pich, Arcadi Santamaria, Phys. Lett. {\bf B424} (1998) 367.
\bibitem{r34} LEP and SLD Working Group, CERN EP/2000-16; LEP and SLD Working Group, presented
by A. Gurtu in a plenary talk at ICHEP 2000, Osaka (2000).
\bibitem{r35} M. Steinhauser, Phys. Lett. {\bf B429} (1998) 158.
\bibitem{r36} BES-II Collaboration, J.Z. Bai et al., Phys. Rev. Lett. {\bf 84} (2000) 594.
\bibitem{r37} CMD-2 Collaboration, R.R. Akhmetshin et al., hep-ex/9904027; Phys. Lett. {\bf B466}
(1999) 392; SND Collaboration M.N. Achasov et al., hep-ex/9809013; Nucl. Phys. {\bf A675} (2000) 391;
Phys. Lett. {\bf B462} (1999) 365.
\bibitem{r38} K.G. Chetyrkin, Phys. Lett. {\bf B390} (1997) 309.
\bibitem{r40} Handbook of Mathematical Functions, Edited by Milton Abramowitz and Irene A. Stegun,
p.556 (ninth edition) Dover (1964).
\end{thebibliography}
\end{document}